\documentstyle[11pt,epsf]{article}
\newcommand{\ewxy}[2]{\setlength{\epsfxsize}{#2}\epsfbox[10 60 640 570]{#1}}

\def\seq {\! = \!}        

\let\ga=\gamma

\let\de=\delta
\let\De=\Delta

\let\eps=\epsilon

\let\la=\lambda
\let\La=\Lambda

\let\del=\nabla
\let\si=\sigma

\let\om=\omega
\let\Om=\Omega

\def\to{\rightarrow}
\let\p=\partial
\let\<=\langle
\let\>=\rangle

\let\txt=\textstyle
\let\dsp=\displaystyle

\let\ad=\dagger
\def\eqn#1{(\ref{#1})}  

\def\beq{\begin{equation}}
\def\eeq{\end{equation}}
\def\ba{\begin{array}}
\def\bea{\begin{eqnarray}}
\def\ea{\end{array}}
\def\eea{\end{eqnarray}}

\def\slash{\!\!\!\!/\,}
\def\sslash{\!\!\!/\,}

\def\comment#1{ \hbox{[{\it Comment suppressed here.}\/]} }
\def\hide#1{}
\def\o{\over}      
\def\O{ {\cal O} }

\def\Tr{\hbox{Tr}}
\def\P{\hbox{P}}

\def\eg{{\it e.g.}}

\def\IR{\relax{\rm I\kern-.18em R}}
\def\IN{\relax{\rm I\kern-.18em N}}
\def\IB{\relax{\rm I\kern-.18em B}}
\def\IE{\relax{\rm I\kern-.18em E}}
\def\ZZ{\relax{\sf Z\kern-.4em Z}}     
\def\IC{{\ontopss{$\scriptscriptstyle \mid$}{{\rm C}}{8.0}{0.415}}}
\def\ontopss#1#2#3#4{\raise#4ex \hbox{#1}\mkern-#3mu {#2}}

\newdimen\pmboffset
\def\oldpmb#1{\setbox0=\hbox{#1}%
 \copy0\kern-\wd0
 \kern\pmboffset\raise 1.732\pmboffset\copy0\kern-\wd0
 \kern\pmboffset\box0} 
\def\pmb#1{\mathchoice{\oldpmb{$\displaystyle#1$}}{\oldpmb{$\textstyle#1$}}
	{\oldpmb{$\scriptstyle#1$}}{\oldpmb{$\scriptscriptstyle#1$}}}

\def\appendix{\par                              
    \setcounter{section}{0}                     
    \setcounter{subsection}{0}
    \renewcommand{\theequation}{\Alph{section}.\arabic{equation}}
    \renewcommand{\thesection}{Appendix \Alph{section}
                \setcounter{equation}{0}  } 
}

\catcode`\@=11

\def\section{
\setcounter{equation}{0}        
\@startsection {section}{1}{\z@}{-3.5ex plus -1ex minus 
 -.2ex}{2.3ex plus .2ex}{\large\bf}}
\renewcommand{\theequation}{\arabic{section}.\arabic{equation}}

\def\subsection{\@startsection{subsection}{2}{\z@}{-3.25ex plus -1ex minus 
 -.2ex}{1.5ex plus .2ex}{\normalsize\bf}}

\def\subsubsection{\@startsection{subsubsection}{3}{\z@}{-3.25ex plus
 -1ex minus -.2ex}{1.5ex plus .2ex}{\normalsize}}

\def\@eqnnum{%
\savebox{\eqnumb}{\rm (\theequation)}%
\settowidth{\numblen}{\usebox{\eqnumb}}%
\makebox[\numblen][l]{\usebox{\eqnumb}~~~\usebox{\eqlabel}}}

\catcode`\@=12

\newcommand{\skipover}[1]{}
\newcommand{\nn}{\nonumber \\}

\def\half {{\txt {1\over 2}}}

\def\Fmn{F_{\mu\nu}}
\def\Fcmn{F_{\! c \, \mu\nu}}
\def\pbar{{\bar p}}
\def\phat{\hat{p}}
\def\psib{{\bar\psi}}
\def\Omb{{\bar\Om}}

\def\={\!=\!}
\def\+{\,+\,}
\def\-{\,-\,}
\def\Lai{\La^{-1}}

\def\delspsl{\pmb{\del}\slash}
\def\delspslc{{\delspsl}_{\! c \,}}
\def\delc{{\del_{\! c \,}}}
\def\delslc{{\del\slash}_{\! c \,}}
\def\Dsl{D\slash}
\def\Dssl{{{\bf D}\slash}}

\newsavebox{\eqlabel}

\catcode`\@=11
\newlength{\numblen}
\newsavebox{\eqnumb}
\def\@eqnnum{%
\savebox{\eqnumb}{\rm (\theequation)}%
\settowidth{\numblen}{\usebox{\eqnumb}}%
\makebox[\numblen][l]{\usebox{\eqnumb}~~~\usebox{\eqlabel}}%
}
\catcode`\@=12

\newenvironment{equationwithlabel}[1]{ %
  \savebox{\eqlabel}{#1}
  \begin{equation}\label{#1} }{\end{equation}\savebox{\eqlabel}{~}}

\newcommand{\beql}[1]{\begin{equationwithlabel}{#1}}
\newcommand{\eeql}{\end{equationwithlabel}}

\newenvironment{eqnarraywithlabel}[1]{ %
  \savebox{\eqlabel}{#1}
  \begin{eqnarray}\label{#1} }{\end{eqnarray}\savebox{\eqlabel}{~}}

\newcommand{\beal}[1]{\begin{eqnarraywithlabel}{#1}}
\newcommand{\eeal}{\end{eqnarraywithlabel}}
\begin{document}

\thispagestyle{empty}

\begin{flushright}
November 1996 \\
FSU-SCRI-96-134 \\
IASSNS-HEP 96/115 \\
\end{flushright}
\vskip 10mm
\begin{center}
{\LARGE \bf Improving Lattice Quark Actions}
\vskip 13mm
{\normalsize M.~Alford}
\vskip 1.3mm
{\small \em School of Natural Sciences\\[0.5mm]
     Institute for Advanced Study\\
     Princeton, NJ 08540}
\vskip 5mm
{\normalsize T.R.~Klassen}
\vskip 1.3mm
{\small \em SCRI\\[0.5mm]
     Florida State University\\
     Tallahassee, FL 32306-4052}
\vskip 5mm

{\normalsize G.P.~Lepage}
\vskip 1.3mm
{\small \em Newman Laboratory of Nuclear Studies\\[0.5mm]
     Cornell University\\
     Ithaca, NY 14853}
\vskip 10mm

{\normalsize \bf Abstract}

\vskip 4mm

\begin{minipage}{4.9in}
We explore the first stage of the Symanzik improvement
program for lattice Dirac fermions, namely the construction of doubler-free,
highly improved classical actions on isotropic as well as anisotropic lattices
(where the temporal lattice spacing, $a_t$, is  smaller than the spatial one).
Using field transformations to
eliminate doublers, we derive the previously presented isotropic D234
action with $\O(a^3)$ errors, as well as anisotropic
D234 actions with $\O(a^4)$ or $\O(a_t^3,a^4)$ errors.
Besides allowing the simulation of heavy quarks within a relativistic
framework, anisotropic lattices 
alleviate potential problems due to unphysical branches of the
quark dispersion relation (which are generic to improved actions),
facilitate studies of lattice thermodynamics, 
and allow accurate mass determinations for particles
with bad signal/noise properties, like glueballs and P-state mesons.
We also show how field transformations can be used to completely
eliminate unphysical branches of the dispersion relation.
Finally, we briefly discuss 
future steps in the improvement program.
\end{minipage}
\end{center}

\newpage
\renewcommand{\thepage}{\arabic{page}}
\setcounter{page}{1}


\section{Introduction}

Lattice calculations suffer from scaling errors, or lattice artifacts, that
typically decrease like some power, $a^n$, of the lattice spacing
(ignoring logarithmic corrections).
Continuum results are obtained as $a\to 0$, but
the cost of a realistic simulation of QCD, for example, grows like
like some large power, $a^{-\omega}$, 
of the inverse lattice spacing
($\om$ is at least~6, but could even be~10 or more~\cite{Gup}).
It is therefore extremely important
to find highly improved actions; they will give accurate results on
much coarser lattices. 

Classical field theory estimates suggest that eliminating 
errors through order $a^2$ and maybe $a^3$   
allows one to model the properties of a
smooth bump with errors of a few percent to a fraction of a 
percent by using a lattice with 3--6 grid points per diameter
of the object in each direction. For hadrons this means that spatial
lattices of spacing $a=0.2-0.4$~fm might suffice for improved actions
(whereas $a=0.05-0.1$~fm are typically used for unimproved actions).
Even considering the computational overhead due to the more complicated
form of improved actions, 
it is clear that being able to work on coarse lattices
would save many orders of magnitude in CPU time.

For pure glue it has already been demonstrated~\cite{Alf1} that this
is possible. In this paper we describe some of the steps that are
necessary to extend these savings to the more difficult problem
of lattice quark actions. Our considerations are mainly classical,
but we will outline where quantum effects seem to play a role and
how to take them into account.
The approach we would like to follow, pioneered by 
Symanzik~\cite{Sym,LW,LWPLB,SW,LPCAC},
is to try to improve lattice actions and fields 
to some finite order in $a$, like $a^2$ or $a^4$. 

For asymptotically free theories, such as QCD,
the terms in the action can be organized by their UV dimensions. 
Symanzik improvement then consists of adding higher dimension
improvement terms to the action, mimicking the effects of the UV
modes left out on the lattice.
To fix the coefficients of these terms, one
then proceeds as follows. Write down all terms with the appropriate
symmetries up to the desired order, with arbitrary coefficients. Tune
the coefficients by matching to a sufficiently large set of
observables, calculated either in perturbation theory or non-perturbatively
in a Monte Carlo simulation. Once the tuning is completed, the
coefficients in the action will be functions of the physical couplings
and a set of redundant couplings (see below).
Improving (composite) field operators involves a similar process, which
must be repeated for each independent field.

The above program is quite difficult to carry through in practice,
not only non-perturbatively, but even in perturbation theory.
 In the past,
standard lattice perturbation theory suffered from the rather
debilitating problem that it did not seem to work very well, at least
compared to continuum perturbation theory. This has now largely been
understood~\cite{LM} as being due to large renormalizations from
tadpole diagrams, which occur in (naive) lattice but not continuum
perturbation theory. 
Using a
simple mean-field type method, known as tadpole improvement,
one can design      more continuum-like operators in which the tadpole
contribution is greatly reduced.  Tadpole improvement has been shown
to work well for a variety of actions 
on surprisingly coarse
lattices~\cite{Alf1,NRQCDccbar,LAT95,LAT96,SCRI,Borici,Bock,Bi}.

We emphasize that tree-level tadpole improvement should be thought of 
as a first step in a systematic procedure of improving lattice actions. 
The next step can be further perturbative improvement, or, if there are
reasons to believe that this is not sufficient, non-perturbative
improvement.

It turns out to be substantially harder to improve lattice fermions
than gluons, 
even  on the classical level. The reason, ultimately,
is the first-order nature of the fermion field equations, which leads
to the well-known doubler problem, which we will discuss later.
For Dirac fermions (quarks),
Wilson~\cite{Wil} solved this problem by adding a second-order
derivative term to the action. This term breaks chiral symmetry at
$\O(a)$. Such errors --- which are much larger than the $\O(a^2)$ errors
of ``naive lattice fermions'' --- are too large for this action to be useful
in coarse lattice simulations.

The point of this work is to present doubler-free quark actions, for light
{\it and} heavy quarks, that are classically improved to high order.
The general tool to construct such actions will be {\it field redefinitions};
they allow one to introduce second-order derivative terms without destroying
improvement.
To allow the simulation of heavy quarks --- and also to avoid potential
problems due to unphysical branches of the quark dispersion relation, which are
generic to improved actions --- we can use 
{\it anisotropic lattices}~\cite{Kar,aniso}. 
Let us discuss these ideas in turn.

Field redefinitions are just changes
of variable in the path integral, so they do not affect spectral quantities
(at least if one takes into account the Jacobian).  Off-shell 
quantities of course do change. Since field redefinitions involve one or more 
free parameters, they lead to so-called {\it redundant} couplings, whose
values can be adjusted at will. This freedom can be used to solve the
doubler problem, for example. In other situations,
in particular on the quantum level, it is very convenient to simplify
an improved action by setting certain couplings to zero. This leads to
the concept of {\it on-shell improvement}, where only spectral quantities
can be obtained directly from the action (by improving composite operators
one can, however, also obtain their matrix elements between physical states).

The ``canonical'' procedure for obtaining a doubler-free quark action
correct up to, say, $\O(a^n)$ classical errors involves the following
three steps:\footnote{The simple recipe to follow is a streamlining
of what can be found in~\cite{SW} together with~\cite{Heatlie}.}

\begin{enumerate}

\item[1.]  Start with the continuum Dirac action and apply a field
redefinition introducing even-order derivative terms into the action.

\item[2.] Expand the continuum operators in the transformed action
in terms of lattice operators up to $\O(a^n)$ errors; this step
will be referred to as the {\it truncation}. The even-order lattice
derivative terms will eliminate the doublers that would be present
without the field redefinition. One can stop here if one is only
interested in spectral quantities; they will be classically correct
up to $\O(a^n)$ errors.

\item[3.] To classically also improve off-shell quantities, {\em undo} the 
field transformation (now on the level of the lattice action). 
The resulting action and fields differ only at $\O(a^n)$ form their
continuum counterparts, and, in contrast to a naive discretization,
have no doublers.

\end{enumerate}


We emphasize that the improved actions so constructed are (classically)
improved in every
respect; the improvement of interactions does not have to be checked
separately.
When applied to lowest order ($n\seq 2$)
the above   procedure gives the Sheikholeslami-Wohlert action,
originally suggested~\cite{SW} as an improvement of the Wilson action.
The next order ($n\seq 4$)
yields the class of ``D234'' actions (in addition to the
second order derivative Wilson term, they also contain third and fourth order
derivative terms).

As alluded to earlier, a problem generic to actions improved beyond $\O(a)$
is the existence of unphysical branches of the free dispersion relation, 
simply due to higher order time derivatives in the action. 
We will refer to these extra branches as lattice {\it ghosts}. 
Their  energies are at the scale of the (temporal) cutoff, so they
will eventually decouple as the lattice spacing is decreased. 
For the lattice spacings used in practice
their effect on, say, the hadron spectrum 
has not been thoroughly studied, but they certainly affect the
renormalization of the improvement terms in the action.
In addition, they can complicate numerical simulations by introducing
oscillations in correlation
functions at small times.

 If either of these issues turns out to be a
problem, one can deal with the ghosts in one of two ways.
\hide{
Either by using field transformations 
to replace the temporal with spatial derivatives. This produces 
somewhat more complicated actions, as we will see, so one might instead
consider using anisotropic lattices with smaller temporal than spatial lattice
spacing,  $a_t<a_s$, to push up the energy of the ghosts and  decouple them. 
}
Firstly, one can use field transformations 
to replace the temporal with spatial derivatives. This produces 
somewhat more complicated actions, as we will see.
Alternatively and secondly,  one therefore might want to
use anisotropic lattices with smaller temporal than spatial lattice
spacing,  $a_t<a_s$, to push up the energy of the ghosts and  decouple them.

Besides effectively solving the potential
problem of ghost branches, the use of anisotropic 
lattices has other advantages:

\begin{enumerate}

\item[$\bullet$] 
By choosing $a_t$ sufficiently small, one can simulate heavy quarks
within a relativistic framework~\cite{LAT96} without the prohibitive cost of a
fine spatial lattice.  This provides a simple 
alternative to the NRQCD~\cite{NRQCD} and Fermilab~\cite{FNAL} formalisms.

\item[$\bullet$] The signal to noise ratio of a correlation function calculated
in a Monte Carlo simulation decays, generically, exponentially in time.
Choosing a smaller $a_t$ gives more time slices with an accurate signal,
allowing for more precise and confident mass determinations.
This is important for particles with bad signal/noise properties, like
P-state mesons~\cite{AKL} and 
glueballs~\cite{MorPea}.

\item[$\bullet$] It facilitates thermodynamic studies --- one of the reasons
being simply that it is easier to take independent derivatives with
respect to volume and temperature if one can vary $a_t$ independent of $a_s$
--- especially at high temperatures.

\item[$\bullet$]  It allows for significant simplifications  in the design of
improved actions.
                    This will be relevant for our D234 actions.

\end{enumerate}

All these advantages come at a price.  Because they have lost part of
their axis-permutation symmetry, anisotropic actions have more
independent coefficients.  This is not a problem at the classical
level, but at the quantum level some of these coefficients will have
to be tuned to restore space-time exchange symmetry.  Large
renormalizations violating space-time exchange symmetry were in fact
seen in first attempts of using anisotropic lattices, see~\cite{Kar}
and references therein. We find that with 
an improved gluon action and
a tadpole improvement prescription
appropriate to anisotropic lattices, such effects are 
relatively small, at 
the level of several percent on coarse lattices~\cite{aniso}.

So far we have concentrated on classical Symanzik improvement;
however,
the improvement of fermion actions is also more difficult on the
quantum level. For pure glue, the largest error at order $a^2$ is the
violation of rotational invariance, which 
can be tuned to zero non-perturbatively, by demanding rotational invariance
of the static potential at physical distances.\footnote{To 
on-shell improve gluons at order $a^2$ on an isotropic
lattice one
has to add two terms to the leading plaquette, which one can choose to
be the ``rectangle'' and the ``parallelogram''~\cite{LW}.  It is a
certain linear combination of the coefficients of these two terms that
can be tuned non-perturbatively by demanding rotational
invariance. Since 
the coefficient of the parallelogram seems to be very small (it certainly 
is to one loop)~\cite{LWPLB,Alf1}, this amounts to an ``almost full''
non-perturbative tuning of the glue at order $a^2$.}
Actually, it seems that most of these errors are removed by tadpole
improvement~\cite{Alf1,aniso}.   

Wilson-type quarks, on the other hand, have $\O(a)$ errors on the quantum
level, and to eliminate them one has to tune a term that violates
chiral but not rotational symmetry. The leading $a^2$ errors 
behave in the opposite way; they violate rotational symmetry
but not chiral symmetry (so they are similar to the errors of gluons).
The $\O(a)$ and (leading) $a^2$ errors of
Wilson-type quarks
therefore have very distinct effects on 
spectral quantities, and can be tuned
iteratively,    even on a non-perturbative level, by demanding chiral
symmetry for the former, and rotational symmetry for the latter.

As for glue it seems that tadpole improvement does quite a good job 
in estimating the coefficient of the $\O(a^2)$ terms that lead to a
restoration of rotational symmetry. Concerning the restoration of chiral
symmetry to eliminate $\O(a)$ quantum errors, L\"uscher et~al have 
recently shown in some beautiful work~\cite{LPCAC} how to implement 
this in practice for the case of SW quarks on Wilson glue.

The outline of the remainder of this paper is as follows. In sect.~2 we discuss
naive lattice fermions, doublers and ghosts. 
We proceed in sect.~3 to describe in more detail the three-step procedure to 
eliminate doublers, which we then apply to derive the Sheikholeslami-Wohlert 
and D234 actions on a general anisotropic lattice. 
Several special cases and variations are also discussed,
including a completely ghost-free D234-like action. 
 In sect.~4 we investigate the large mass behavior of the D234 actions. 
We conclude in sect.~5 with a brief summary and 
sketch of future steps in the improvement program.

Appendices~A and~B summarize our notation for euclidean continuum and
lattice QCD, respectively. 
The reader might want to skim these appendices before starting with the
main text, and later refer back to them as necessary.
Appendix~C discusses
the dispersion relation of the D234 actions. Finally, in appendix~D we give some
formulas useful in the tadpole improvement of the D234 actions.

Brief accounts of various 
parts of this work have appeared earlier in~\cite{LAT95,LAT96,Melb,Schl}.


\section{Naive Lattice Fermions, Doublers, and Ghosts}\label{sec:Naive}

Discussions of Dirac fermions on the 
lattice\footnote{It is even more difficult to put {\em chiral} fermions on the 
lattice~\cite{NiNi}. This we will not attempt. 
See~\cite{NN} for recent work on this problem.}
usually start with the so-called {\em naive} lattice fermions, 
specified by the fermion 
operator $\del\slash + m$. Here $\del_\mu$ is the usual
first order, anti-hermitean, covariant lattice derivative,
\beq
 \del_\mu \psi(x) ~\equiv~
 {1\over 2a_\mu}\, \biggl[ U_\mu(x) \psi(x+\mu) - U_{-\mu}(x) \psi(x-\mu)\biggr] 
\eeq
in terms of the link field $U_\mu(x)$ (cf.~appendix~B for details).
This derivative differs at order $a_\mu^2$ from the continuum Dirac operator
$D_\mu = \p_\mu -i A_\mu$.

One way of stating the origin of the doubler problem is that $\del_\mu$
{\it decouples even and odd} sites of the lattice. This leads to a doubling
(per direction) of the number of degrees of freedom on the lattice.
If it were not for this problem, 
 naive fermions would provide a lattice discretization of Dirac
fermions with order $a^2$ errors. Similarly, one could use an 
improved operator,  such as
\beq
\delc_\mu ~\equiv~ \del_\mu \, 
                  \Bigl(1 \- {1\over 6} \, a^2_\mu \Delta_\mu\Bigr )
          ~=~ D_\mu + \O(a_\mu^4) ~,
\eeq
where $\De_\mu$ is the standard second order lattice derivative of appendix~B,
and the subscript ``$c$'' stands for ``continuum-like''. The fermion operator
$\delslc + m$ defines what we will refer to as {\it naive improved}
lattice fermions. They would provide a lattice action with only order $a^4$
errors --- again, if we could ignore the doublers.

The simplest way of 
discussing the doubler problem for a generic lattice action is in
momentum space.
Let us consider the dispersion relation $E=E({\bf p})$, obtained
from the poles of the free euclidean propagator, $p_0 = p_0({\bf p})$,
via $E=\pm i p_0$. The two signs correspond to particle and anti-particle.
For simplicity we will factor out the sign and consider only solutions
where the (real part of the) energy is positive. The quantitative details
of the dispersion relations of  the actions considered in this paper are
discussed  in appendix~C. In figure~\ref{fig:NaiveNaiveI} we show the massless
dispersion relations of naive and naive improved fermions on an isotropic 
lattice. There are several noteworthy features.

\begin{figure}[tb]
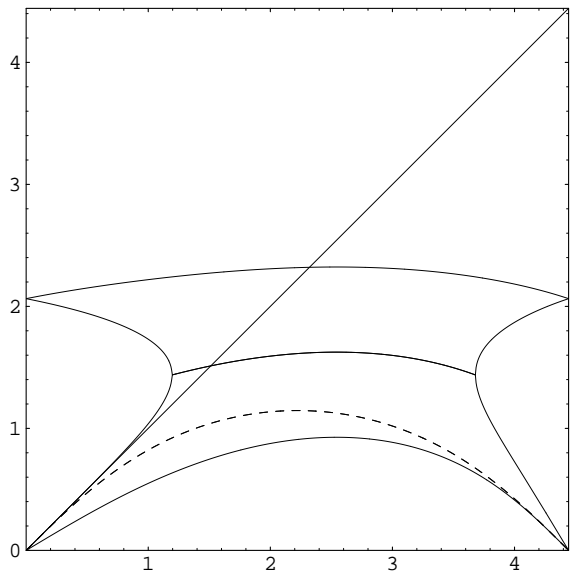

\vskip 12mm
\centerline{\ewxy{Epp_NaiveNaiveI_mc0_1_A1.epsi}{100mm}}
\vskip -12mm
\caption{
(Real part of the) massless dispersion relation $a E= a E({\bf p})$ as a function 
of $a |{\bf p}|$, with ${\bf p} \propto (1,1,0)$, for naive fermions (dashed) 
and naive improved fermions (solid) on an isotropic lattice.
For comparison we also show the dispersion relation of continuum fermions 
(thin solid).
}
\label{fig:NaiveNaiveI}
\vskip 2mm
\end{figure}

\begin{enumerate}

\item[$\bullet$]
For naive fermions the one branch of the dispersion relation presented in 
figure~\ref{fig:NaiveNaiveI} is purely real. Since we can only exhibit a cross
section of the energy surface, one sees only one of the {\em spatial}
doublers, which account for half of the doublers.
The term ``spatial doubler'' refers to the fact that for each possible energy
$E$ there are generically eight momenta ${\bf p}$ (with all components positive)
such that $E=E({\bf p})$.

\item[$\bullet$]
We also can not show that for each possible $E=E({\bf p})$ of naive fermions 
there is another pole of the propagator at $E + i\pi/a_0$. This (as well as 
the existence of the spatial doublers) follows from the fact
that in momentum space the action of naive fermions only depends on
$a_\mu \bar{p}_\mu \equiv \sin(a_\mu p_\mu)$, which is invariant under
$a_\mu p_\mu \rightarrow  \pi - a_\mu p_\mu$. 
These complex poles constitute the 
{\em temporal} doublers.

\item[$\bullet$]
For naive improved fermions the picture is more complex. There are now four
branches. 
The lowest branch
is somewhat pathological in that its imaginary part is $\pi/a$ for all
momenta and in that its real part is {\em lower} than that of the 
physical branch.
It is easy to see that this branch is related to the temporal doubler of the 
naive fermion action.

\end{enumerate}

Clearly, neither of these two actions corresponds to what one might expect a 
lattice Dirac fermion to look like. Due to the doublers, the naive fermion 
action actually describes 16 Dirac fermion species in the continuum 
limit~\cite{KarSmit}. In addition to spatial doublers, naive improved fermions
have {\it ghosts} (or {\it lattice ghosts}, if confusion could arise),
as extra branches of the dispersion relation will be called from now 
on.\footnote{To be sure, these ghost branches are not related to the ghosts
 appearing in loop
diagrams of perturbation theory in non-abelian gauge theories. These branches
do not describe independent particles; 
they are just lattice artifacts related to the lattice ``particle'' described
by the physical branch of the dispersion relation.
What justifies naming them ghosts is that they
usually (but not always) 
give negative contributions to the spectral representation
of correlation functions. In practice this leads to a characteristic ``dip''
in effective mass plots.}
As mentioned in the introduction, if the ghosts should turn out to be
a problem, there are always ways of eliminating them, to be discussed
in the next section.

A solution to the doubler problem
 was proposed by
Wilson~\cite{Wil},  who suggested to add a second-order derivative term 
(now known as the Wilson term) to give the fermion operator
\beq
M_{{\rm W}} ~=~ m_0 \+ \sum_\mu \, \biggl(
      \ga_\mu \del_\mu - {1\over 2} r a \Delta_\mu \biggr ) ~,  
\eeq
on an isotropic lattice. $r$ is the so-called Wilson parameter.
It is easily shown (cf.~sect.~\ref{sec:Impr}) that whereas there
is one ghost branch for generic $r$, there is none for $r\seq 1$.
There is no doubler problem for any $r>0$.

The Wilson action with $r\seq 1$ therefore solves the doubler problem without
introducing any ghosts. However, the addition of the Wilson term introduces
$\O(a)$ errors, which are too large for this action to be useful on coarse
lattices.
Sheikholeslami and Wohlert~\cite{SW} described a modification of the
Wilson action that has only $\O(a^2)$ errors for on-shell quantities.
Their action differs from the Wilson action  by a 
$\si \! \cdot \! F \equiv \sum_{\mu\nu}\si_{\mu\nu} F_{\mu\nu}$ 
term (cf.~appendices A and B
for notation), commonly known as the {\it clover} term.
 The free dispersion relation of this
action is therefore identical to that of the Wilson action.

In~\cite{SW} only on-shell improvement was considered. Later it was 
realized~\cite{Heatlie} that by performing a suitable change of variables on the
fields in the Sheikholeslami-Wohlert action, one can also calculate off-shell
quantities up to $\O(a^2)$ errors, at tree level.
We will present a succinct derivation of all this, and its generalization to
higher orders of improvement on anisotropic lattices, in the next section.


\section{Improved Lattice Fermion Actions}\label{sec:Impr}

\subsection{Improvement without Doublers}\label{sec:Impr1}

As seen in the previous section, the naive and naive improved
fermion actions have a doubler problem. More generally, this is true
for any fermion matrix of the form
\beq
 \sum_\mu ~\ga_\mu \del_\mu ~\bigl ( 1 \- 
  b_\mu \, a_\mu^2 \, \De_\mu \+ d_\mu \, a_\mu^4 \, \De_\mu^2 
                                       \+ \ldots \bigr ) ~,
\eeq        
which preserves chiral symmetry. 
(And even more generally, this follows from the Nielsen-Ninomiya 
theorem~\cite{NiNi}; see also~\cite{KarSmit}.)  
To avoid doublers we therefore should, following
Wilson, introduce chiral symmetry breaking,
even-derivative terms $\De_\mu$ (or powers thereof)
into the action. However, we would like to avoid the $\O(a)$ errors
that a naive addition of a Wilson term entails. This can be achieved
by a {\it field redefinition}. 

The simplest way to proceed is to perform
the field redefinition in the continuum and only subsequently discretize
the action.
Starting with the continuum action
\beq
 \int \, \psib_c \, M_c \, \psi_c ~\equiv~
      \int d^4x ~ \psib_c(x) \, (D\slash + m_c)\, \psi_c(x) ~,
\eeq
we perform a field redefinition
\bea\label{cv}
\psi_c               &=& \Om_c ~\psi   \nn
\psib_c              &=& \psib ~\Omb_c \nn
\psib_c ~M_c ~\psi_c &=& \psib ~M_\Om ~\psi ~, 
 \qquad M_\Om ~\equiv~ \Omb_c ~M_c ~\Om_c ~. 
\eea
Note that a field transformation does not affect spectral quantities, at
least if we take into account the Jacobian of the transformation.
Classically the Jacobian does not matter. 
On the quantum level its leading effect is to renormalize the gauge
coupling.

Our canonical choice of field redefinition is
(with $\Omb_c$ acting to the right)   
\beq\label{cvstd}
\Omb_c ~=~ \Om_c ~,  \quad 
 \Omb_c \, \Om_c ~=~ 1 \- {r a_0 \over 2} \, (D\slash - m_c) ~.
\eeq
At this point $a_0$ is just a constant with the dimension of length,
but in the subsequent lattice discretization $a_0$ will become the 
temporal lattice spacing.

The transformed fermion operator $M_\Om$  reads 
\bea\label{MOm}
 M_\Om &=& m_c \+ D\slash  \- {1\over 2} \, r a_0 \, ( D\slash^2 - m_c^2) \nn
       &=& m_c (1 + \half r a_0 m_c) \+ D\slash
       - {1\over 2} \, r a_0 \,
           \Bigl(\sum_\mu D_\mu^2 \+ \half \si \! \cdot \! F \Bigr) ~,
\eea
where we used eq.~\eqn{DDexpns}.
We can {\it now} put the above action on the lattice by
discretizing $D\slash$ and $D_\mu^2$ and $\Fmn$ to some order, $a^n$,
using, for example, eqs.~\eqn{pexpn}, (B.10), \eqn{cloverrep} 
and~\eqn{impcloverrep} in the $n\seq 4$ case. Let us call the lattice action
so obtained $M$.

If one is only interested in spectral quantities, one can use the
propagator $G = M^{-1}$ in further calculations. Off-shell quantities will 
then generically have $\O(a)$ errors, since $\Om_c \seq 1 + \O(a)$
and therefore  $\psi \seq \psi_c + \O(a)$.
However, as our third step,
we can also improve off-shell quantities by {\em undoing} the field 
transformation.
To do so, we use the obvious lattice versions of the operators
in eq.~\eqn{cvstd}, which differ from them at order $a^n$. Let us call these
operators $\Om$ and $\Omb$. 
The action obtained by undoing the change of variable is
$\Omb^{-1} \, M \, \Om^{-1}$, using fields that differ from the
original continuum fields only at $\O(a^n)$.
The propagator of this action is
\bea\label{corrprop}
 G ~=~ \Om ~M^{-1} ~\Omb ~=~ M_c^{-1} + \O(a^n) ~. 
\eea

Note that undoing the field transformation on the lattice does
not lead to the (re)appearance of doublers.

\subsection{The Sheikholeslami-Wohlert  Action and $\O(a)$ Terms}\label{sec:SW}

Using the leading discretization of the derivatives 
in~\eqn{MOm} gives
\bea\label{MSW}
 M_{{\rm SW}} ~=~  
                  m_c (1 + \half r a_0 m_c) \+ \del\slash 
 \- {1\over 2} r a_0 ~\sum_{\mu} \Delta_\mu   
 \- {1\over 4} r a_0 \, \si \! \cdot \!  F  ~.
\eea
This is the Sheikholeslami-Wohlert (SW) action on an anisotropic 
lattice.
For $\Fmn$ one can use the so-called {\it clover} representation 
(cf.~appendix~B), which has $\O(a^2)$ errors.
By construction this action has classical $\O(a^2)$ errors for spectral 
quantities; also for off-shell quantities if we undo the change of
variable. To obtain the Wilson action one must by hand set the 
clover term, $\sigma \! \cdot \! F$, to zero,
thereby incuring an $\O(a_0)$ error
in the presence of a non-trivial gauge field.

\begin{figure}[tb]
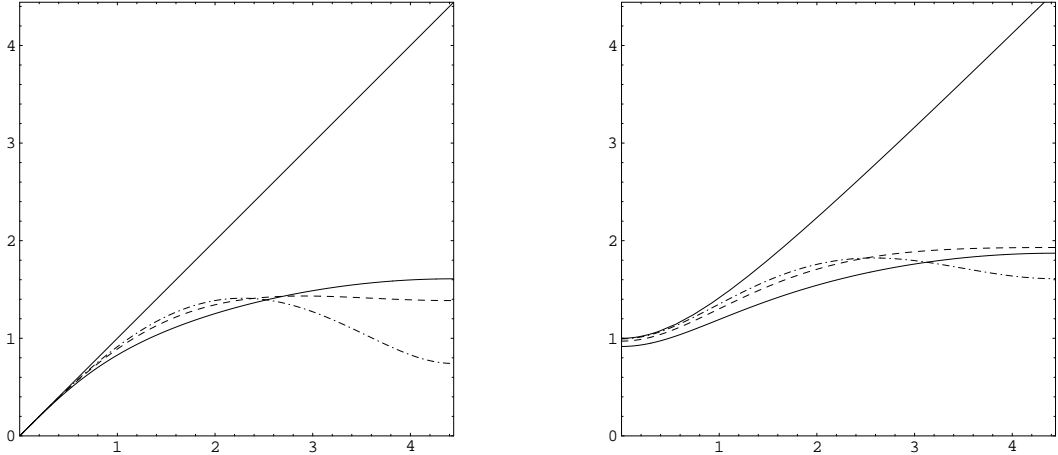

\vskip 12mm
\mbox{ \ewxy{Epp_Wil_mc0_125_A1.epsi}{80mm} 
       \ewxy{Epp_Wil_mc1_125_A1.epsi}{80mm} } 
\vskip -12mm
\caption{
The energy $a_s E({\bf p})$ of free SW/Wilson fermions with $r\seq 1$ 
as a function of $a_s |{\bf p}|$ with ${\bf p} \propto (1,1,0)$. 
On the left we show the massless case on 1:1 (solid), 2:1 (dashed), 
and 5:1 (dot-dashed) lattices, as well as continuum fermions (thin solid). 
On the right we show the same for mass $a_s m_c=1$.
}
\label{fig:Wil}
\vskip 2mm
\end{figure}

This action has no doublers for any $r>0$; 
it has no ghost branches
either if and only if $r\seq 1$, which is therefore the canonical choice.
For future reference we show in figure~\ref{fig:Wil}
the free dispersion relations for the SW/Wilson action with $r\seq 1$ 
for various anisotropies $a_s/a_t$ and masses.

Recall that in general more operators are needed for quantum improvement
(even on-shell) than for classical improvement. For Wilson-type quark
actions on an isotropic lattice, however, it is easy to see that the
clover and Wilson terms are the only ones allowed at $\O(a)$ by gauge
and (discrete) rotational symmetry~\cite{SW}. Their coefficients of course
renormalize on the quantum level.
By a field transformation of the canonical form discussed above
one can adjust the coefficient of one of these terms to any desired value.
It is natural do so for the Wilson term; in the SW case to maintain its 
``bare'' coefficient at the canonical value $r\seq 1$, for example.
To eliminate quantum $\O(a)$ errors one then has to tune the coefficient
of the clover term.

On an anisotropic lattice the situation is slightly more complicated.
The allowed operators at $\O(a)$ consist of the spatial and temporal
parts of the Wilson and clover terms, and the additional operator
$[\ga_0 D_0 , \sum_i \ga_i D_i ]$.
The most general field transformations $\Om_c$ and
$\Omb_c$ allowed in this situation lead to three redundant operators, so
that one has to tune two coefficients at $\O(a)$. These can be choosen to
be the spatial and temporal parts of the clover term.
Note that on an anisotropic lattice one must also allow a relative
coefficient between the temporal and spatial kinetic terms at $\O(a^0)$,
which can be tuned non-perturbatively by demanding a relativistic dispersion
relation for the pion, say, at small masses and momenta.

\subsection{The D234 Actions}

Going to the next order in the expansion of the continuum derivatives
in~\eqn{MOm} gives the class of D234 actions
\bea\label{MDiiii}
 M_{{\rm D234}} &=&  m_c (1 + \half r a_0 m_c) \+ \sum_\mu\,
      \ga_\mu \del_\mu \, ( 1 - b_\mu a_\mu^2 \De_\mu ) \nn
&& \- {1\over 2} \, r a_0  \,
    \Bigl( \sum_\mu \Delta_\mu  \, \+ \, \half \, \si \! \cdot \! F \Bigr)
            \+ \sum_\mu c_\mu a_\mu^3 \De_\mu^2 
\eea
where, at this point,
\beq\label{bcdefDc}
  b_\mu ~=~ {1\over 6} \, , \qquad\quad  c_\mu ~=~ {r a_0\over 24 a_\mu}
\eeq

The specific D234 action defined by the coefficients in~\eqn{bcdefDc}
will be referred to as ``D234c($r$)'', where ``c'' refers to the
fact that this action is obtained by our ``canonical'' field redefinition
without any further modifications.
If we use an improved representation of the field strength, as in
eq.~\eqn{impcloverrep}, this action only has $\O(a^4)$ classical
errors. There is no canonical choice of $r$ for this action.
It will generically have three ghost
branches, as illustrated in fig.~\ref{fig:Diiiic} for the case of 
$r\seq 1$ (for $r\seq 2$ there are only two ghosts, but the lowest one
is too low for this choice to be interesting, except perhaps on very
anisotropic lattices).

\begin{figure}[htb]
\vskip 12mm
\centerline{\ewxy{Epp_D234c1_mc0_12_A1.epsi}{100mm}}
\vskip -12mm
\caption{
As in figure~\protect\ref{fig:Wil}, for the
massless D234c(1) action on 1:1 (solid) and    
2:1 (dashed) lattices. We only show the real part of the energy,
relevant for the top branch of each anisotropy, which has imaginary
part $\pi/a_0$, and to the right of the branch point on the 1:1 lattice.
}
\label{fig:Diiiic}
\vskip 2mm
\end{figure}

It remains to be investigated what effect these 
ghost branches have on the quantum level, but one should certainly consider
designing actions with fewer and higher-lying ghost branches.
As we will see, the necessary ``tuning'' of the D234 actions introduces
classical errors in addition to the $\O(a^4)$ ones. However, this is
probably irrelevant, since the latter errors are unlikely
to dominate on the quantum level anyhow.

\subsection{Tuning the D234 Actions}

To investigate the ghosts let us study the free dispersion relation 
corresponding to the general D234 action~\eqn{MDiiii}. 
The details are discussed in appendix~C, to which we refer for the proof
of any non-obvious facts we will use.
For generic $r, b_\mu$ and $c_\mu$ the 
dispersion relation is a quartic equation for 
 $\sinh(a_0 E/2)^2$, $E=E({\bf p})$,  so there will be
three ghost branches.\footnote{Remember that we do not count the particle
anti-particle symmetry $E\leftrightarrow -E$ in the number of solutions.}
Since the coefficients of the quartic are real, the energies
will be real, come in complex conjugate pairs, or have imaginary part
$\pi/a_0$.  Note that the qualitative branch
structure ({\it e.g.}~the number of branches) depends only on the temporal
coefficients $r, b_0$ and $c_0$. For $m_c\seq 0$ and ${\bf p=0}$ the only way
a lattice spacing enters the dispersion relation is via $a_0 E$. This implies
that for small momenta and masses the height of the ghosts is inversely
proportional to the temporal lattice spacing.

The most basic question we can ask about the ghosts, is how many ghost 
branches  we can completely eliminate by a
suitable choice of the free parameters. We summarize the conclusions of 
appendix~C concerning this question  as follows:

\begin{enumerate}

\item[$\bullet$]
If we choose $b_0 \seq 2 c_0$  there will be at most two ghosts.

\item[$\bullet$]
If we further choose  $r = 1 - 2 b_0$~ or  $b_0 = 0$ there is (at most) one ghost.

\item[$\bullet$]
The only way to eliminate all ghosts is to choose 
$r\seq 1, b_0 \seq c_0 \seq 0$,
which is of course the standard SW/Wilson case (if $r\neq 1$ the SW/Wilson
action will have one ghost). 

\end{enumerate}

If we want to go beyond the SW action we will therefore have at least
one ghost branch. Let us now discuss our ``favorite'' D234 actions
on isotropic and anisotropic lattices in turn.

\subsubsection{Isotropic D234 Action}\label{sec:Wii}

On an isotropic lattice one will presumably prefer an isotropic action
with manifest space-time exchange symmetry, so that one does not
have to restore this symmetry by tuning on the quantum level.
With $b_\mu={1\o 6}$, the above results imply that when we require 
one ghost we must choose $r={2\o 3}$ and $c_\mu={1\o 12}$. We will
refer to this action as the ``isotropic D234'' action. It was 
introduced in~\cite{LAT95}.

\begin{figure}[htbp]
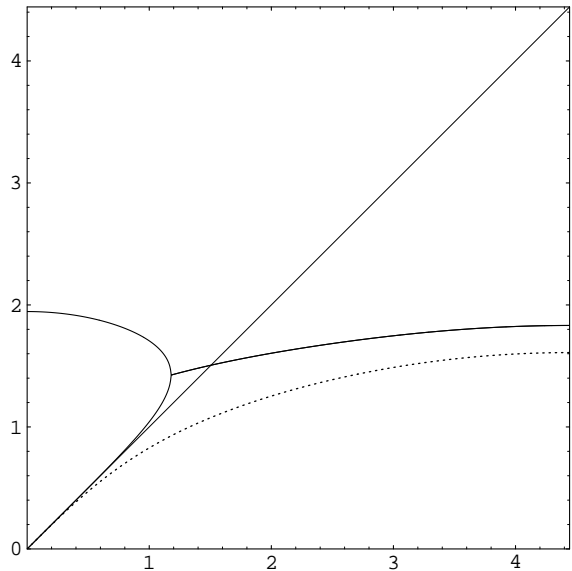

\vskip 12mm
\centerline{\ewxy{Epp_D234_Wil_mc0_1_A1.epsi}{100mm}}
\vskip -12mm
\caption{
As in figure~\protect\ref{fig:Diiiic}, for the
massless isotropic D234 (solid) and 
SW (dotted) actions. 
}
\label{fig:Wii}
\vskip 2mm
\end{figure}

Since the $c_\mu$ violate eq.~\eqn{bcdefDc} this action has
classical $\O(a^3)$ errors. At zero mass and momentum the one ghost branch 
of this action is at $a E(0)\seq \ln 7 \approx 1.9459$.
Its massless dispersion relation is shown
in figure~\ref{fig:Wii}, together with that of the SW/Wilson action.

We will see in sect.~\ref{sec:W} that, like the SW and Wilson actions, 
this action can be coded very efficiently using the ``projection
trick''.

\subsubsection{Anisotropic D234 Action}\label{sec:aniso}

\begin{figure}[htbp]
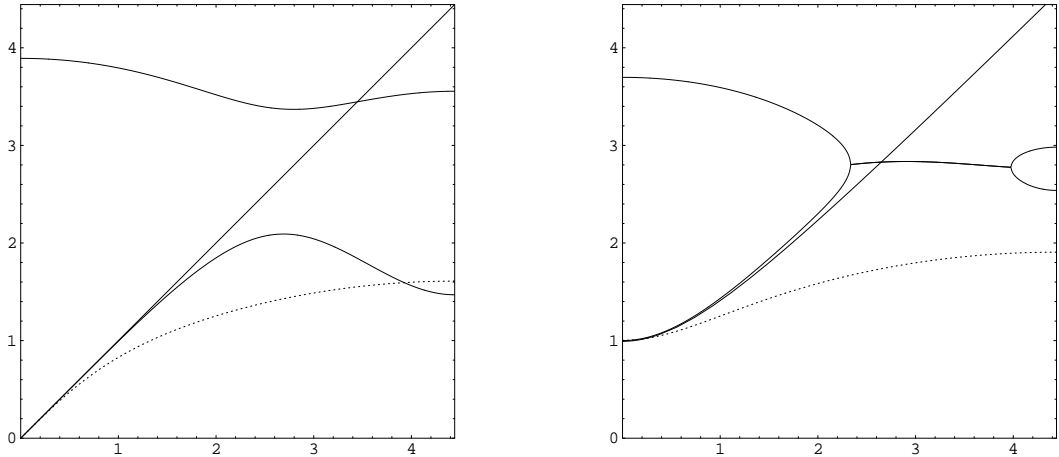

\vskip 12mm
\mbox{ \ewxy{Epp_D234isa2_Wila1_mc0_A1.epsi}{80mm} 
       \ewxy{Epp_D234isa2_Wila1_m1_A1.epsi}{80mm} } 
\vskip -12mm
\caption{
As in fig.~\protect\ref{fig:Wil}, for D234i(${2\over 3}$)  on a
2:1 (solid) and SW/Wilson fermions on a 1:1 (dotted) lattice.
The real part of $E({\bf p})$ is shown to the right of the
D234i(${2\over 3}$) branch point.
In the massive case the bare masses are tuned so that $a_s E(0)\seq 1$.
}
\label{fig:Dis}
\vskip 2mm
\end{figure}

On an anisotropic lattice we can have one ghost with only
$\O(a_0^3, a_\mu^4)$ errors simply by modifying the coefficient $c_0$
of the D234c(${2\o 3}$) action to be $c_0={1\o 12}$. We will refer 
to this action by the name D234i(${2\o 3}$).\footnote{The ``i'' stands
for mass-``independent'', since the coefficients in this action enjoy this
property. In the next subsection we will describe a closely related D234
action with mass-dependent coefficients, to which we have 
previously~\cite{LAT96} given the name D234(${2\o 3}$).}
Its dispersion relation on a 2:1 lattice is compared in 
figure~\ref{fig:Dis} with that of the SW action.
Note in particular the impressive dispersion relation in the massive
case, indicating that such an action might be very useful for heavy
quark simulations.

We should point out
that the restriction of the D234i(${2\o 3}$) action
to an isotropic lattice does {\it not} give the isotropic D234 action.
Because the spatial $c_i$ of the former were chosen to not
introduce $a_i^3$ errors, this action has anisotropic coefficients even 
on an isotropic lattice; it was designed for use on anisotropic lattices.

\subsubsection{Variations}\label{sec:var}

By relaxing the requirement of just one ghost one can construct
actions that might be interesting for sufficiently anisotropic
lattices. We will not discuss these here, but just make the
general remark that for larger anisotropies $a_s/a_t$ 
it is advantageous to choose larger values of
$r$. Otherwise one will recover spatial doublers, as is obvious
from the fact that our canonical field transformation~\eqn{cvstd}
$\Om_c \to 1$ as $a_0 \to 0$ for fixed $r$.\footnote{If one
avoids the reappearance of doublers by letting $r a_0$ approach
some non-zero limit as $a_0 \to 0$, one will obtain one ghost
branch, with energy $2/r a_0$ for small masses and momenta.}

We should briefly discuss one modification of the D234i(${2\o 3}$)
action. Namely, it is possible to construct a very similar D234 action
that in the free case has only $\O(a^4)$ classical errors, with $\O(a_0^3)$
errors entering only in the presence of a gauge field.
This action is obtained by a somewhat more complicated change of
variable from the continuum Dirac action,
\bea\label{prodcvopDiiii}
 \Omb_c ~=~ \Om_c ~, ~~~
 \Omb_c ~\Om_c ~=~ 1 - {1\over 2} r a_0 ~(D\slash - m_c - K_{\pmb{\eps}}) ~,
    ~~~ K_{\pmb{\eps}} = \sum_\mu \eps_\mu a_\mu^2 \ga_\mu D_\mu^3 
\eea
with free parameters $\eps_\mu$ in addition to $r$.
Requiring one ghost and $\O(a_0^3)$ errors only in the presence
of a gauge field leads uniquely to the choice of coefficients
\bea\label{magicbc}
\eps_0 &=& {1\over 6} ~{1 \over 1 + {3\over 4} a_0 m_c} \nn
   b_0 &=& {1\over 6} ~{1+{1\over 4} a_0 m_c \over 1+ {7\over 12} a_0 m_c} \nn
   c_0 &=& {1\over 2} ~b_0 \nn
   r   &=& {2\over 3} ~{1+{3\over 4} a_0 m_c \over 1+ {7\over 12} a_0 m_c} \, ,
\eea
as well as $\eps_i=0$, $b_i={1\o 6}$ and $c_i = r a_0/24 a_i$.
We will refer to this action as D234(${2\o 3}$)
(labelling it by the value of $r$ at $m_c\seq 0$).
This was the action used in the
anisotropic lattice simulations described in~\cite{LAT96}.
Note that numerically this action differs significantly from
D234i(${2\o 3}$) only for very large masses; for $m_c\seq 0$ they
are identical.

Finally, we remark that it is easy to improve the dispersion relation
of D234-like actions at large momenta still further by introducing
suitable fifth and sixth order derivative terms.  This can be done
by a field transformation and/or by simply adding such terms to the action.
However, in the relevant momentum regime the hadron dispersion relations
measured in simulations of D234 actions are already so 
good, even on coarse lattices~\cite{LAT95,LAT96},
that the additional cost and complication from the higher derivative 
terms seems unjustified.

\subsection{Other Actions}

We conclude this section by briefly discussing several other classes
of actions (or other ways of writing actions) that are of interest
for various conceptual and practical reasons.

\subsubsection{Ghost-free D234-like Actions}\label{sec:gf}

We will now demonstrate that it is straightforward to write down
a highly improved action, at tree level, that has no ghosts whatsoever.
This comes at a price, of course. Such an action is more complicated and
therefore more costly to simulate.

The idea is to use field transformations to
eliminate the cubic temporal terms $\ga_0 \del_0 \De_0$ in the naive
improved fermion action in favor of spatial terms.
Starting with the continuum action
$M_c = m_c + \Dsl$, we perform a field transformation
$\psi_c = \Om_{c1} \, \psi, \psib_c = \psib \, \Omb_{c1}$ with\footnote{This
field transformation is similar to ones used in~\cite{FNAL}.}
\bea\label{cvopiDiiiigf}
 \Om_{c1} &=&  1 \+ {a_0^2\over 12} \, \Bigl[
       D_0^2 - (\Dsl - m_c)(\Dssl + m_c ) \Bigr] \nn
 \Omb_{c1} &=&   1 \+ {a_0^2\over 12} \, \Bigl[
       D_0^2 - (\Dssl + m_c )(\Dsl - m_c) \Bigr] \, ,
\eea
where the purely spatial derivative $\Dssl = \sum_i \ga_i D_i$.
This gives
\beq
\ba{rcl}
\Omb_{c1} \, M_c \, \Om_{c1} &=&
  m_c \+ \Dsl + {1\over 6}a_0^2\ga_0D_0^3 \+ \delta K_c \+ \O(a^4),\\[2ex]
  \de K_c &=& \dsp {a_0^2\over 12} \{m_c^2 -\sum_i D_i^2 -\half \si\cdot F \,
  , \,  \Dssl + m_c\}
\ea
\eeq
Note that, when discretized, the term $\Dsl + {1\over 6}a_0^2\ga_0D_0^3$
will not contain any lattice time derivative above the first, and there is
no other temporal derivative in the action.
We can now proceed with a second change of variable to introduce even
derivatives, defined by $\Omb_{c2} = \Om_{c2}$ and 
\beq
 \Omb_{c2} \, \Om_{c2} \, = \, 1 \+
  \half r a_0 \, 
     \Bigl(\Dsl + {1\over 6}a_0^2\ga_0D_0^3 - m_c -\delta K_c \Bigr) \, .
\eeq
This implies
\bea
 \Omb_{c2} \, \Omb_{c1} \, M_c \, \Om_{c1} \, \Om_{c2} &=& m_c 
 (1+\half r a_0 m_c) \+
   \Dsl + {1\over 6}a_0^2\ga_0D_0^3 \+ (1+r a_0 m_c)\, \delta K_c \nn
 && \- {1\over 2} r a_0 \, ( \Dsl + {1\over 6}a_0^2\ga_0D_0^3 )^2 \+ \O(a^4) \,
.
\eea
Finally, we discretize this action. 
Using
\beq\label{deltplusdelspcsq}
 ( \Dsl + {1\over 6}a_0^2\ga_0D_0^3 )^2 \, = \, \sum_\mu \De_\mu
   \- {1\over 12} \, \sum_i a_i^2 \De_i^2 \+ {1\over 2} \si \cdot F
   \+ \O(a_0^2,a^4) \, ,
\eeq
we find
\bea
 M_{{\rm D234gf}} &=&  m_c  (1+\half r a_0 m_c) \+ {\del\slash}_0 \+
 \delspslc \+ (1+r a_0 m_c)\, \delta K \nn
 && \- {1\over 2} r a_0 \,
  \Bigl(\sum_\mu \De_\mu + \half \si \cdot F \Bigr)
    \+ {r a_0\over 24} \, \sum_i a_i^2 \De_i^2 \,\, , \\
\de K &=& {a_0^2\over 12} \,
     \{m_c^2 - \sum_i \De_i -\half \si\cdot F \,
       , \, \delspsl + m_c \} = \de K_c + \O(a^4) \, ,
\eea
where ${\del\slash}_0= \ga_0\del_0$ is the unimproved temporal lattice
derivative, $\delspslc$ is the improved spatial derivative (see
\eqn{delcdef}) and for the field strength $F$ one should use an
improved discretization.  This action has $\O(a_0^3,a^4)$ errors in on-shell
quantities.
As usual, by undoing the above field transformations one can achieve the same
errors for off-shell quantities.

\begin{figure}[tb]
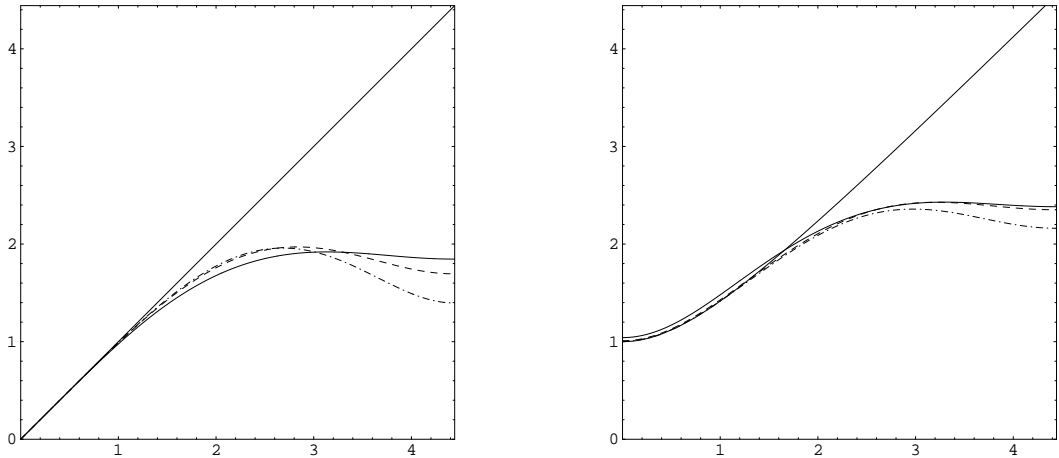

\vskip 11mm
\mbox{ \ewxy{Epp_D2346gf_mc0_123_A1.epsi}{80mm}
       \ewxy{Epp_D2346gf_mc1_123_A1.epsi}{80mm} }
\vskip -11mm
\caption{
As in figure~\protect\ref{fig:Wil}, for the ghost-free D234-like action
on 1:1 (solid), 2:1 (dashed) and 3:1 (dot-dashed) lattices.
}
\label{fig:Diiiigf}
\vskip 2mm
\end{figure}

Obviously the above action is the analog of the D234 actions of the previous
sections, with $\delta K$ appearing in place of cubic and quartic
temporal derivative terms.
Now, however, we have just one branch if we set $r \= 1$, as for the Wilson
and SW actions. This is illustrated in figure~\ref{fig:Diiiigf}.

For the $a_0^3$ errors to be negligible compared to the $a^4$ ones, we
can again use anisotropic lattices. In that case the ghost branches of
the D234 actions of the previous section are presumably harmless, and
it seems doubtful that having no ghosts outweighs the disadvantage of
having to include the costly anti-commutator term $\delta K$.
Comparisons to simulations with this action should, however,
allow one to discern whether the ghost branches have any effect
besides that on correlation functions at small times.
                                                       
Another, less ambitious ghost-free action can easily be constructed if 
one is willing to tolerate $a_0^2$ in addition to $a_0^3$ and $a^4$ errors.
Such an action can be obtained, for example, by simply neglecting the 
temporal third and fourth order terms in the D234c(1) action.

\subsubsection{D34 Action}\label{sec:Diii}

\def\D34fig{
\begin{figure}[tbp]
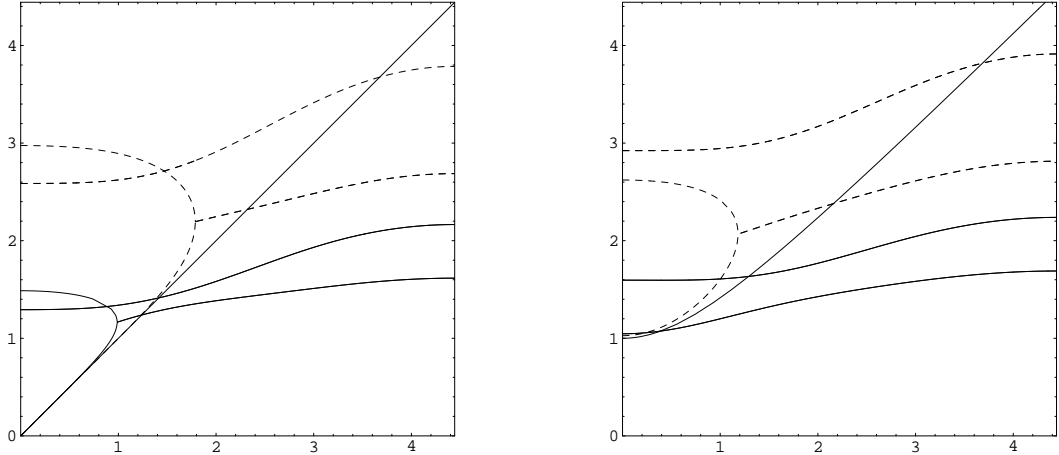

\vskip 12mm
\mbox{ \ewxy{Epp_D34_mc0_12_A1.epsi}{80mm} 
       \ewxy{Epp_D34_mc1_12_A1.epsi}{80mm} } 
\vskip -12mm
\caption{
As in figure~\protect\ref{fig:Dis},
for D34 fermions with $c_\mu = {1\o 6}$ on 1:1 (solid) and 2:1 
(dashed) lattices. Note that this action has four branches.
They consist of two pairs of complex conjugate solutions for sufficiently
large momenta.  For small momenta there are usually two real branches.
For $a_s m_c \seq 1$ on an isotropic lattice, however, all branches
are complex even at small momenta.
}
\label{fig:Diii}
\vskip 2mm
\end{figure}
}

\D34fig

Recently a tadpole-improved version of
 an improved action discussed in~\cite{EguKaw} was used in a Monte Carlo
simulation of the hadron spectrum~\cite{FW}. This action has third and fourth
order derivative terms, but no second order Wilson term. We will therefore refer
to it as the D34 action. Generalized to an anisotropic lattice it reads
in our notation
\beq
M_{{\rm D34}} = m_c \+ \delslc \+
                   \sum_\mu c_\mu a_\mu^3 \De_\mu^2 \, .
\eeq
In~\cite{FW} an isotropic lattice with all $c_\mu \seq {1\over 6}$ was used.
The D34 action is obtained from the naive improved fermion action
not by a change of
variable, but simply by adding the $\De^2$ term, which leads to an $a^3$
error (and means that the fields are already improved up to $a^3$ errors; 
no change of variable
has to be undone). This action is of course a special case of the general class
of D234 actions, with $r=0$ and $b_\mu = {1\over 6}$.
It has three ghosts, except for $c_0 = {1\o 12}$, when there are two.
There is no obvious canonical choice for the $c_\mu$; for all values of 
$c_\mu$ the ghost branches seem to lie rather low. In figure~\ref{fig:Diii}
we show various dispersion relations for the $c_\mu \seq {1\over 6}$ case.

\subsubsection{W-Actions}\label{sec:W}

Recall that the Wilson and SW actions with $r\seq 1$ can be efficiently
coded using the ``projection trick'', which exploits the fact that the
``Wilson operator''~\cite{John}
\beq
 W_\mu ~\equiv~ \del_\mu - {a_\mu \over 2} \ga_\mu \De_\mu
\eeq
can be expressed in terms of (spinor-) projection operators. Most
conveniently this is expressed as
\beq
 \ga_\mu W_\mu ~=~ -\del_\mu^{+} P_\mu^{-} + \del_\mu^{-} P_\mu^{+} \; ,
\eeq
where $P_\mu^{\pm} \equiv \half (1\pm \ga_\mu)$ are
projectors on two-dimensional, orthogonal subspaces (for fixed $\mu$), and
\beq
\del_\mu^{\pm} \psi(x) ~\equiv~ \pm{1\over a_\mu} \biggl(
 U_{\pm \mu}(x) \psi(x\pm \mu) - \psi(x) \biggr) 
\eeq
are forward and backward derivatives.

Actions that can be expressed as (low-order) polynomials in the $W_\mu$,
\beq
 M ~=~ \sum_\mu \, {1\over a_\mu} \, Q_\mu(a_\mu \ga_\mu W_\mu)
\eeq
(with the possible addition of mass and 
clover      terms, cf.~below) will be
cheap to code. Another advantage of such an action is that one can immediately
read off the number of branches of the dispersion relation: If 
$Q_0(x) = x + \ldots$ is an $n$-th order polynomial, there are exactly $n$
branches,  counting possible degeneracies (the proof of this statement is left
as an exercise to the reader).
 Note that if all $Q_\mu(x)$ are $n$-th order polynomials, the
class of such ``W23{\ldots}$n$'' actions is a proper subset of the class of
``D23{\ldots}2$n$'' actions considered previously, which generically have
$2n$ branches.

It is interesting to ask when such W-actions are improved. We first remark that
the continuum derivative can be written as
\beq
a_\mu \ga_\mu D_\mu ~=~ -\ln(1-a_\mu \ga_\mu W_\mu) ~=~
      a_\mu \ga_\mu W_\mu + {1\over 2} (a_\mu \ga_\mu W_\mu)^2
 + {1\over 3} (a_\mu \ga_\mu W_\mu)^3 + \ldots
\eeq
Truncation of this expansion does, however, not seem to lead to promising
(on- and off-shell) improved actions, since the expansion converges too slowly
and the ghost branches lie too low.

Alternatively, we may ask when a W-action is only on-shell improved, related 
to some order in $a$ by our canonical field transformation~\eqn{cvstd}
to the continuum Dirac action.
Restricting ourselves to isotropic actions
from now on --- the improved actions in question are
not simply expressed in terms of $W_\mu$ on anisotropic lattices --- we write
\beq
M ~=~ m_c (1+\half r a m_c) \+ {1\over a}\, \sum_\mu \, Q(a\ga_\mu W_\mu) \-
      {1\over 4} \, r a \, \si\! \cdot \! F \; .
\eeq
For a first-order polynomial, $Q(x) = x$, we recover, of course, the
SW action for $r\seq 1$, with $\O(a^2)$ errors. 
To see which higher order 
polynomials correspond to improved actions,
we expand out the $W_\mu$ in terms of $\del_\mu$ and $\De_\mu$ and compare
with the results of the previous subsections (specifically, the D234c($r$)
actions and their higher order analogs).

The second order polynomial $Q(x) = x + {1\over 6} x^2$ with $r\seq {2\o 3}$ 
gives a ``W2'' action, which has $O(a^3)$ errors. 
It is identical to the isotropic D234 action~\cite{LAT95} of 
sect.~\ref{sec:Wii}.
This implies that the application of the isotropic D234 operator is
only about twice as expensive as that of the $r\seq 1$ SW or Wilson 
operator.

For a third order polynomial one can reduce the errors to $\O(a^4)$ by choosing
$Q(x) = x +{5\over 22} x^2 + {2\over 33} x^3$ and $r={6\o 11}$ to give a ``W23''
action (one must now also use an improved $F$ in the clover term).
This action is equal to a D23456 action with
$r={6\o 11}$, $b_\mu = {1\o 6}$, $c_\mu = {1\o 44}$,  $d_\mu = {2\o 33}$,  
$e_\mu = {1\o 33}$, 
where $d_\mu$ and $e_\mu$ are the coefficients of fifth, respectively,
sixth order derivative terms, defined analogously to $b_\mu$ and $c_\mu$.
This action has two ghost branches; for all masses and momenta they  
form a complex-conjugate pair. For zero mass and momentum
 their energies are $a E(0) \approx 1.528 \pm 0.897 \, i$.


\section{Large Mass Behavior of the D234 Actions}\label{sec:largem}

Our D234 actions were initially designed with light quarks in mind. In
contrast to~\cite{FNAL}, for example, where renormalization conditions are
imposed on mass-shell, even for heavy quarks, we effectively do so at zero
quark mass, since we always expand in powers of $a_\mu$, assuming $a_\mu p_\mu$
to be small. Heavy onium systems are non-relativistic, so masses are large
but momenta are small. One would therefore expect that by going to an anisotropic
lattice, where $a_t m_c$ is small, our D234 actions could be used to simulate
such systems on lattices where the spatial lattice spacing is still quite large.
Decreasing the temporal lattice spacing by a factor of, say, $3-5$, 
would present a
relatively minor increase in cost, given the exciting prospect of obtaining
accurate results for {\it e.g.}~charmonium within a relativistic framework. 
The charmonium system is 
difficult to simulate. It is light enough for 
the NRQCD expansion to become problematic, 
but too heavy for the usual light
quark actions on isotropic lattices to be accurate. A highly improved
quark action on an anisotropic lattice seems taylor-made to simulate such
a system. 

To get a quantitative idea of how small we have to choose $a_t$, let us
investigate in the free case  when the D234 actions break down at large masses.
As indicators of break-down we will consider $E(0)/m_c$ and the 
``effective velocity of light'' $c({\bf p})$, defined by
\beq\label{ciidef}
 c({\bf p})^2 ~=~ {E({\bf p})^2 - E(0)^2 \over {\bf p }^2 } ~,
\eeq
where $E({\bf p})$ is meant to be the physical branch of the energy.
We know that for small masses $E(0)/m_c$  and $c(0)$ are 1 up to order 
$\O(a_t^3 m_c^3)$ or  $\O(a_t^4 m_c^4)$ corrections.  
For large masses, it is clear from the figures
presented in sect.~\ref{sec:Impr}, 
that for the relevant D234 actions the two lowest real 
branches of the energy (at fixed momentum) will eventually merge.
\hide{
For zero momentum the branch point occurs at the same value of $a_t m_c$ for
any anisotropy. For non-zero momentum this becomes true only asymptotically,
for large anisotropy. 
}
(For zero momentum this branch point occurs at the same value of $a_t m_c$ for
any anisotropy. For non-zero momentum this becomes true only asymptotically,
for large anisotropy.)
At the branch point $c({\bf p})$ diverges.

\hide{ 
\begin{table}[tp] \centering
\vskip 6mm
\begin{tabular}{ | l | c | c | l | c | }
\hline
Action &  ~~$a_s {\bf p}$~~  & ~~$\xi$~~  &  ~~~$a_s m_c^\star$  & 
     $E({\bf p})/\sqrt{m_c^{\star 2} + {\bf p}^2}$\\ \hline
D234(${2\over 3}$) & $(0,0,0)$& $n$ &  ~1.333333 $\times \, n$  & 1.207078\\
            & $(1,0,0)$&  1  &  ~0.824907  & 1.196987\\
            & $(1,0,0)$&  2  &  ~2.458000  & 1.204840\\
            & $(1,0,0)$&  3  &  ~3.864431  & 1.206099\\
            & $(1,1,0)$&  1  &  ~0.188486  & 1.061800\\
            & $(1,1,0)$&  2  &  ~2.226849  & 1.202456\\
            & $(1,1,0)$&  3  &  ~3.723256  & 1.205096\\
\hline
D234(1)     & $(0,0,0)$& $n$ &  ~1.369955 $\times \, n$ & 1.210034\\
            & $(1,0,0)$&  1  &  ~0.882199  & 1.201366\\
            & $(1,0,0)$&  2  &  ~2.537112  & 1.208142\\
            & $(1,0,0)$&  3  &  ~3.977947  & 1.209207\\
            & $(1,1,0)$&  1  &  ~0.204830  & 1.077965\\
            & $(1,1,0)$&  2  &  ~2.313639  & 1.206115\\
            & $(1,1,0)$&  3  &  ~3.840835  & 1.208358\\
\hline
D234(2)     & $(0,0,0)$& $n$ &  ~1.465104 $\times \, n$ & 1.216485\\
            & $(1,0,0)$&  1  &  ~0.978663  & 1.183111\\
            & $(1,0,0)$&  2  &  ~2.729722  & 1.209680\\
            & $(1,0,0)$&  3  &  ~4.264987  & 1.213538\\
            & $(1,1,0)$&  1  &  ~0.112278  & 1.046132\\
            & $(1,1,0)$&  2  &  ~2.508105  & 1.202133\\
            & $(1,1,0)$&  3  &  ~4.129387  & 1.210469\\
\hline
\end{tabular}      
\vskip 2mm
\caption{$m_c^\star$ is the value of $m_c$ at which the two lowest 
         branches of the dispersion relation merge, at a fixed momentum.
         We give $m_c^\star$ for several D234 actions, momenta, and anisotropies.
         We also show $E({\bf p})/\protect\sqrt{m_c^{\star 2} + {\bf p}^2}$, 
         which is a measure
         of how much the dispersion relation deviates from the continuum one
         at the branch point. Note that for the D234(${2\over 3}$) action the 
         zero momentum branch 
         point occurs at mass $a_t m_c = a_t m_c^\star = {4\over 3}$, 
         with $a_t E(0) = \ln 5$.
}
\label{tab:mcbp}
\vskip 4mm
\end{table}
}

\begin{figure}[tbp]
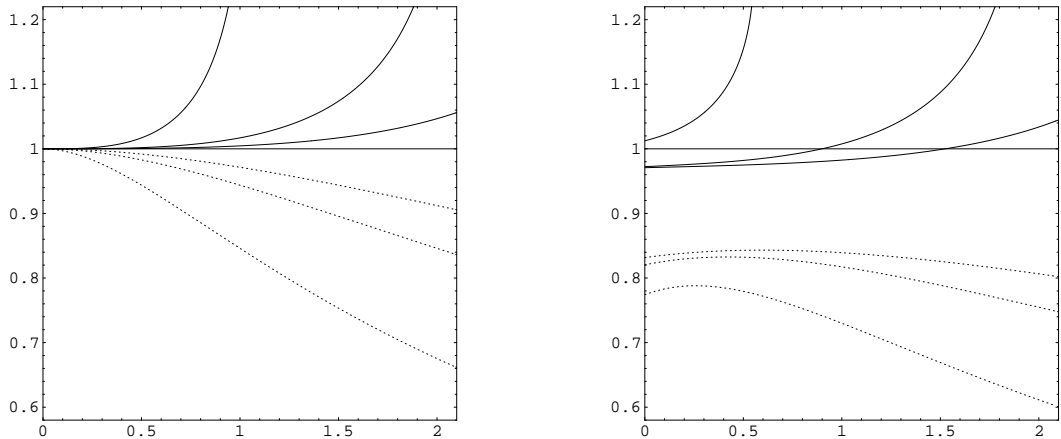

\vskip 11mm
\mbox{ \ewxy{c0.001_D234is_123_A1.epsi}{80mm}
       \ewxy{c1_D234is_123_A1.epsi}{80mm} }
\vskip -11mm
\caption{
~$c({\bf p} \! = \! {\bf 0})\,$ (left) and 
~$c({\bf p} \! = \! (1,0,0)/a_s)\,$ (right)
as functions of $a_s m_c$ for D234i(${2\over 3}$) 
(solid) and SW/Wilson (dotted) fermions on 1:1, 2:1 and 3:1 lattices.
}
\label{fig:c0}
\vskip 2mm
\end{figure}

In figure~\ref{fig:c0}  we compare  $c({\bf p})$ for the 
D234i(${2\over 3}$) and SW/Wilson fermions for various anisotropies.
We show $c({\bf p})$ as a function of mass for both ${\bf p = 0}$ and the
(quite large) value ${\bf p} = (1,0,0)/a_s$.

As expected, $c({\bf p})$ stays close to 1 for a larger and larger mass range
as the anisotropy increases. Note how much better the D234 action behaves in this
respect than the SW/Wilson action, in particular at non-zero momentum.
It is also interesting to observe that for the SW/Wilson action 
$c({\bf p})$ decreases more or less monotonically as a function of both
$m_c$ and $|{\bf p}|$, whereas for the D234 action  $c({\bf p})$  increases
as a function of $m_c$ for fixed ${\bf p}$,  but (more or less) decreases as a 
function of $|{\bf p}|$ for fixed $m_c$ (an isotropic lattice provides
an exception to this latter statement).
This is presumably one of the reasons we find in Monte Carlo
simulations that the D234
actions exhibit an excellent dispersion relation for various hadrons up to
surprisingly large masses and momenta.

As is clear from figure~\ref{fig:c0}, for the D234 actions the transition from 
$c({\bf p}) \approx 1$ to divergence is quite rapid. This provides a sensitive
indicator of the breakdown of the D234 actions at large masses.
When $c({\bf p})$ begins to get larger than 1, the D234 actions should not be
trusted anymore. When this happens, it should only be necessary to decrease
$a_t$ by a relatively modest amount to be able to simulate the quark mass
of interest.

To reinforce this last point, let us mention that 
among the anisotropic D234 actions we discussed, the finite-mass errors
are largest for the D234i(${2\o 3}$) action, where
\bea\label{leadingEcexpn}
%
 {E(0)\over m_c} &=& 1 \+ {1\over 18} \, a_t^3 m_c^3  \+ \O( a_t^4 m_c^4 ) \\
      c(0)       &=& 1 \+ {1\over  9} \, a_t^3 m_c^3  \+ \O( a_t^4 m_c^4 ) ~.
\eea

For comparison, the corresponding expansions for the free SW/Wilson action
are
\bea
 {E(0)\over m_c} &=& 1 \- {1\over 6} \, a_t^2 m_c^2  \+ \O( a_t^3 m_c^3 ) \nn
      c(0)       &=& 1 \- {1\over 3} \, a_t^2 m_c^2  \+ \O( a_t^3 m_c^3 ) ~.
\eea

We see that not only do the D234 actions have a larger power in their scaling
errors, they are also blessed with small coefficients.
The coefficients in the above expansions for various other 
actions can be found in appendix~C.  There we also discuss the error that 
arises in the presence of
a non-zero gauge field. Again we find that the coefficients of the error terms
are very small as long as $a_t m_c \leq \O(1)$.

For the large mass error in $c(0)$ to stay below the one or two percent level,
one must choose $a_t m_c < 0.5$ for the D234 actions.
For the SW action one must satisfy the much tighter constraint $a_t m_c < 0.2$
to achieve the same error. Given this bound for the D234 actions,
we can attain a meson mass
\bea
   m_{q\bar{q}} ~\approx~ 2 m_c ~=~ {2\xi\over a_s} \, a_t m_c 
\eea
in the charmonium range ($m_{q\bar{q}} \approx 3.0\,{\rm GeV}$) 
with a spatial lattice
spacing of $a_s^{-1} \approx 600-1000\,{\rm MeV}$ and anisotropy $\xi = 3-5$.
For the SW action the lattice spacings would have to be almost three times
smaller with the same anisotropy, making a Monte Carlo simulation roughly
two orders of magnitude more expensive. 
Needless to say, on an isotropic lattice a simulation of charmonium
in a relativistic formalism would be more expensive by many, many 
additional orders of magnitude.



\section{Discussion and Conclusions}\label{sec:Disc}

We have seen that
using field transformations it is easy to design highly improved, doubler-free
quark actions at tree level.
We emphasize that the improvement includes all interactions between
quarks and gluons, not just the free quark dispersion relation.
We have discussed several actions, representing different
compromises between the conflicting aims of high level of improvement,
absence of ghost branches, and simplicity.
The actions we are currently using in simulations are the
 isotropic D234 (sect.~3.4.1) and
the anisotropic D234i(${2\o 3}$) (sect~3.4.2) actions.
However, given the ingredients provided, the reader can concoct 
many more flavors of actions.

The         actions we constructed are just moderately more expensive to 
simulate
than the SW action. For example, the isotropic D234 action is only about 
twice, the generic D234 action about four times as expensive than the
isotropic SW action with $r\seq 1$ (a slight additional overhead might
be incured in preconditioning).

The next task, of course, is to put some quantum flesh on these classical
bones.
Exploratory quenched simulation results using tadpole improvement
have appeared~\cite{LAT95,LAT96}, and are very encouraging. Further 
work is in progress~\cite{AKL}.
In terms of the general improvement program, the next step is to check
if a non-perturbative tuning of the low-order coefficients in the 
action is necessary. To eliminate all errors up to $\O(a)$, one would
only have to tune the clover term, for an isotropic action. It was
recently shown~\cite{LPCAC} how to implement this in practice, by
demanding the restoration of chiral symmetry at zero quark mass.
In~\cite{LPCAC} the case of SW quarks on Wilson glue was considered,
but the methods apply equally well to other actions.

As mentioned in the introduction (cf.~sect.~\ref{sec:SW} for more details), 
for anisotropic actions there are more coefficients 
that can renormalize on the quantum level, and
in principle have to be tuned in order to restore 
space-time exchange symmetry.       
We hope that suitable tadpole improvement on anisotropic lattices~\cite{aniso}
will reduce these renormalizations so that their effect on physical
quantities is only on the few percent level, where
they can be neglected; but this is an issue to be decided by empirical study.

Already at $\O(a^0)$ one might have to introduce a ``bare velocity of
light'' into the action, to restore space-time exchange symmetry at
leading order.  The required renormalization is easily determined
non-perturbatively in this case, by measuring the dispersion relation
of a pion, say, at small masses and momenta. For on-shell improvement
at $\O(a)$ there is one additional coefficient compared to the
isotropic case, which one can choose to be the relative coefficient of
the temporal and spatial parts of the clover term.  Using the methods
of~\cite{LPCAC} with a suitable background field, for example, it
should be possible to determine this renormalization. Finally, we 
should point out that on an anisotropic lattice the effect of these
renormalizations is suppressed, since the overall coefficient of the
$\O(a)$ terms is smaller by a factor $a_t/a$ compared to the isotropic 
case (for fixed $r$).

In summary, the D234 actions show great promise for accurate QCD simulations
on coarse lattices of improved glue, for both light and heavy quarks. The 
low-order terms in the actions require further study,
and, if necessary, should be determined non-perturbatively.
We hope that, as for improved glue, quark actions with 
negligible
$\O(a)$ and only small
quantum $\O(a^2)$ errors will give accurate results for {\it all} observables
on coarse lattices. The next task of our program will be the determination
of renormalization constants like $Z_A$ and $Z_V$ for the D234 actions. Once
quenched QCD is satisfactorily understood, it is time for simulations of
full QCD. 
Realistic simulations of full QCD should finally be possible
on coarse lattices.

\vskip 9mm
\noindent
{\bf Acknowledgements}

\noindent
We would like to thank Robert Edwards, Urs Heller, Tony Kennedy,
Paul Mackenzie, Stefan Sint, and John Sloan for discussions.
This work is
supported by DOE grants DE-FG05-85ER25000, DE-FG05-92ER40742 and
DE-FG02-90ER40542. 

\newpage

\appendix

\section{Euclidean Continuum QCD}\label{app:contqcd}

We write the action of euclidean SU($N$) gauge theory in four dimensions as
\beq
 S_g[A] \, = \, {1\over 2 g^2} \int d^4x ~\Tr~ F_{\mu\nu}(x) F_{\mu\nu}(x) ~,
\eeq
where $F_{\mu\nu}(x)$ is the su($N$)-valued hermitean field strength. In our
conventions the covariant derivative is written in terms of the hermitean
gauge field $A_\mu(x)$ as
\beq
 D_\mu ~=~ \p_\mu - i A_\mu
\eeq
so that
\beq
 F_{\mu\nu} ~=~ i [D_\mu,D_\nu] ~=~ \p_\mu A_\nu -\p_\nu A_\mu -i 
[A_\mu,A_\nu] ~.
\eeq
In terms of traceless hermitean su($N$) generators $T_a$, $a \seq 1,\ldots,N^2-1$,
normalized by $\Tr(T_a T_b) = {1\over 2}\de_{ab}$, we write
\beq
  F_{\mu\nu}(x) = F_{\mu\nu}^a(x) T_a, ~~~~~
  A_{\mu}(x)    = A_{\mu}^a(x) T_a ~.
\eeq

The parallel transporter from a point $x$ to $y$ along a curve ${\cal C}_{yx}$
is
\beq\label{partrans}
 U[{\cal C}_{yx}] ~=~ 
 \P \exp\biggl( i \int_{ {\cal C}_{yx} } dx'_\mu A_\mu(x') \biggr) \, \, 
 \in {\rm SU(}N{\rm )} ~,
\eeq
where P denotes path ordering, stipulating that in a series or product 
expansion of the exponential the fields at a point earlier on the curve are to
be placed to the right of fields at later points. Under a local gauge 
transformation, $\La(.) \in$ SU($N$),
\bea\label{gtFU}
  F_{\mu\nu}(x) &\to& \La(x) \, F_{\mu\nu}(x) \, {\La^{-1}}(x) \nn
  U[{\cal C}_{yx}] &\to& \La(y) \, U[{\cal C}_{yx}] \, {\La^{-1}}(x) ~.
\eea
Recall that the inhomogenous transformation law of $A_\mu(x)$ has been
{\em designed} so that the parallel transporter transforms covariantly as above.

\vskip 3mm 
The action of a Dirac fermion coupled to a gauge field is
\beq
 S_f[\psi,\psib] ~=~ \int d^4x ~\psib(x) ~(D\slash + m) ~\psi(x) ~,
\eeq
where $D\slash = \sum_\mu \ga_\mu D_\mu$ in terms of the euclidean gamma matrices
defined by
\bea
\ga_\mu &=& \ga_\mu^\ad \nn
\{ \ga_\mu, \ga_\nu \} &=& 2 \de_{\mu\nu} ~.
\eea
The spinor fields $\psi(x)$ and $\psib$ also carry a suppressed color index
in the vector representation of SU($N$). Under a gauge transformation
\bea
  \psi(x) &\to& \La(x) \, \psi(x) \nn
  \psib(x) &\to&  \psib(x) \, \Lai(x)
\eea
and $D_\mu \psi(x)$ transforms like $\psi(x)$ (again, by construction of 
$A_\mu(x)$).

We will need some identities involving the hermitean 
$\si_{\mu\nu}$ matrices defined by
\beq
\si_{\mu\nu} = -{i\over 2}[ \ga_\mu,\ga_\nu ] = -\si_{\nu\mu} \, ,
\qquad {\rm or} \qquad \ga_\mu \ga_\nu = \de_{\mu\nu} + i \si_{\mu\nu} \, .
\eeq
This implies 
\beq\label{DDexpns}
{1\over 2} \{ \ga_\mu D_\mu, \ga_\nu D_\nu \} ~=~ \de_{\mu\nu} D_\mu^2 + 
                                {1\over 2} \si_{\mu\nu} F_{\mu\nu} \nn
\eeq

\section{Lattice QCD}\label{app:latqcd}

We will consider a  four-dimensional euclidean hypercubic lattice of extent
$L_\mu$ and lattice spacing $a_\mu$ in direction $\mu=0,1,2,3$. Since we will
always work in euclidean space (except when we consider dispersion
relations) the use of $\mu \seq 0$ to denote the euclidean time direction
should not cause confusion.

Points are labelled by $x,y,\ldots$,  as in the continuum.
 When all spatial lattice spacings are identical,
they are denoted by $a_s$; the temporal lattice spacing will then be
denoted by $a_t \seq a_0$.
We will refer to $\xi\equiv a_s/a_t$ as the {\em anisotropy} of the lattice,
and sometimes call such a lattice a ``$\xi:1$ lattice''. For an isotropic lattice
we set $a\equiv a_\mu$.

We use $i, j, \ldots$ for spatial indices, and boldface letters for spatial
vectors.  The notation $x\pm\mu$ is a shorthand
for $x\pm a_\mu \hat{\mu}$, where $\hat{\mu}$ is a unit vector in the positive
$\mu$-direction.

When working on anisotropic lattices one should in principle be very careful
in specifying the lattice spacing errors of various quantities. To avoid
cumbersome notation, we will be
careful only to distinguish errors of the form $\O(a_\mu^n)$ from 
$\O(a^n)$ errors, where the latter denotes any errors that are not of the
form  $\O(a_\mu^n)$ or are a sum of such terms with different $\mu$.  

Gauge field dynamics on a lattice is expressed in terms of the link field
$U_\mu(x)$, which takes values in the gauge group. $U_\mu(x)$ is a parallel
transporter from $x+\mu$ to  $x$, along a straight line.
In terms of an underlying continuum gauge field $A_\mu(x)$ we therefore have
(cf.~appendix~A)
\beq
 U_\mu(x) ~=~ \P \exp\biggl( -i \int_x^{x+\mu} dx'_\mu A_\mu(x') \biggr) ~,
\eeq
where now (because we changed the sign of the path compared to 
eq.~\eqn{partrans})
fields at points {\it later} on the path are to be placed to the right of 
earlier fields.	
We will employ  the notation $U_{-\mu}(x) \equiv U_\mu(x-\mu)^\dagger$ for the
parallel transporter from $x-\mu$ to $x$.

Using the link field $U_\mu(x)$ is the only known way of maintaining exact
gauge invariance on the lattice  
(not preserving manifest gauge invariance would necessitate a 
costly tuning of the various gauge couplings).
Under a gauge transformation
\beq
 U_{\pm \mu}(x) ~\to~ \La(x) \, U_\mu(x) \, \Lai(x\pm \mu) \, .
\eeq

With the help of the link field it is trival to construct gauge-covariant first- 
and second-order lattice derivatives via
\bea
 \del_\mu \psi(x) &\equiv& 
 {1\over 2a_\mu}\, \biggl[ U_\mu(x) \psi(x+\mu) - U_{-\mu}(x) 
           \psi(x-\mu)\biggr] \\ 
 \De_\mu \psi(x)  &\equiv& 
 {1\over a_\mu^2} \, \biggl[ U_\mu(x) \psi(x+\mu) + U_{-\mu}(x) \psi(x-\mu)
                                       -2 \psi(x) \biggr] \, .
\eea
The operators $\del_\mu$ and $\De_\mu$  are the building blocks of our
lattice fermion actions. 

It is useful to observe that in momentum space (with $U_\mu \equiv 1$)
they correspond to
\bea\label{pbarphat}
 \del_\mu &\leftrightarrow& \, \, i \pbar_\mu  \, , \, \, \, \, \,
                      a_\mu \pbar_\mu \equiv \sin(a_\mu p_\mu) \\
 \De_\mu &\leftrightarrow& - \phat_\mu^2  \, , \, \, \, \, 
                      a_\mu \phat_\mu \equiv 2 \sin(a_\mu p_\mu/2)
\eea
Momenta on a lattice can be taken to lie in the Brillouin zone, defined
by $-\pi/a_\mu < p_\mu \leq \pi/a_\mu$.
Note that $\pbar_\mu$, in contrast to $\phat_\mu$, has a ``doubler problem'',
{\it i.e.}~it vanishes at the edge of the Brillouin zone $a_\mu p_\mu = \pi$.

Since $\del_\mu$ and $\De_\mu$ are gauge-covariant, many identities involving
these operators are easy to prove by going to a gauge in which $U_\mu \equiv 1$,
where such identities reduce to relations between $\pbar_\mu$ and $\phat_\mu$.
For example, 
\beq\label{delDeCom}
  \del_\mu \De_\mu ~=~ \De_\mu \del_\mu ~,
\eeq
and
\beq\label{deldelexpn}
 \del_\mu \del_\mu ~=~ \De_\mu + {1\over 4} a_\mu^2 \De_\mu \De_\mu \, .
\eeq
Other identities we will 
need       follow from
\bea\label{pexpn}
 p_\mu &=& \pbar_\mu \Bigl( 1 \+ {a_\mu^2\over 6} \, \phat_\mu^2 
\+ {a_\mu^4\over 30}\, \phat_\mu^4 \+ {a_\mu^6\over 140}\, \phat_\mu^6 
  \+ \ldots  \Bigr)  \\
 p_\mu^2 &=& \phat_\mu^2 \+ {a_\mu^2\over 12} \, \phat_\mu^4 \+
     {a_\mu^4\over 90} \, \phat_\mu^6 \+ {a_\mu^6\over 560} \, \phat_\mu^8 \
          + \ldots
\eea

The operator $\del_\mu$ has $\O(a_\mu^2)$ errors compared to the continuum
derivative,
\beq
 \del_\mu ~=~ D_\mu + \O(a_\mu^2) ~.
\eeq
Eq.~\eqn{pexpn} implies that a more continuum-like first-order covariant 
derivative, involving only next-nearest neighbor sites, can be defined as 
follows
\beq\label{delcdef}
 \delc_{\mu} ~\equiv~ \del_\mu \Bigl( 1 - {1\over 6} a_\mu^2 \De_\mu \Bigr)
           ~=~ D_\mu + \O(a_\mu^4) \, .
\eeq

The clover operator is a well-known lattice representation of the
field strength  that agrees with the
continuum $\Fmn$ up to $\O(a^2)$ errors. To define it, let us expand on
the path notation used for the parallel transporter in~appendix~A.
Namely, let us introduce a shorthand for lattice paths,
defined recursively so that ${\cal C}_{yx}(\mu)$ denotes the path from 
$x+\mu$ to $x$ to $y$, and $x\seq {\cal C}_{xx}$.  Path ordering implies that
$U[{\cal C}_{yx}(\mu)] = U[{\cal C}_{yx}] U_\mu(x)$. Of course, 
$U[{\cal C}_{xx}] \seq 1$.

The clover operator           we use,  $\Fmn^{(cl)}(x)$,
can be expressed in terms of the sum of the link fields around the four 
plaquettes  bordering on $x$ (all counter-clockwise, say),  as 
\bea\label{cloverdef}
 \Fmn^{(cl)}(x) &\equiv&  {1\over 4 a_\mu a_\nu} {\cal T}\biggl(
 U[x(\mu)(\nu)(-\mu)(-\nu)] + U[x(\nu)(-\mu)(-\nu)(\mu)] + \nn
& & \qquad\qquad\,\,  U[x(-\mu)(-\nu)(\mu)(\nu)] + U[x(-\nu)(\mu)(\nu)(-\mu)]
 \biggr) \, ,
\eea
where ${\cal T}(M)$ is the (color-)traceless imaginary part of an
$N\times N$ matrix,
\beq
 {\cal T}(M) ~\equiv~ {1\over 2}(M-M^\dagger) \- 
                              {i\over N} \, {\rm Im} \Tr \, M \; .
\eeq
A simple calculation shows that
\beq\label{cloverrep}
 \Fmn^{(cl)}(x) ~=~
  \Fmn(x) \,+\, {1\over 6}\,( a_\mu^2 D_\mu^2 \,+\, a_\nu^2 D_\nu^2 )\,\Fmn(x) 
   \,+\, \O(a^4) \, .
\eeq
\hide{
We can therefore define a more continuum-like lattice field strength via
\bea\label{impcloverrep}
 \Fcmn(x) &\equiv&  \Fmn^{(cl)}(x) \, - \, {1\over 6} \, 
         ( a_\mu^2 \De_\mu \,+\, a_\nu^2 \De_\nu ) \, \Fmn^{(cl)}(x) \nn
 &=& {5\over 3}  \, \Fmn^{(cl)}(x) \; - \; {1\over 6} \, \biggl[
    U_\mu(x) \Fmn^{(cl)}(x+\mu)U_{-\mu}(x+\mu) \+ \nn
& & \qquad\qquad\qquad\quad \, \,
 U_{-\mu}(x) \Fmn^{(cl)}(x-\mu)U_{\mu}(x-\mu)  \-
 (\mu\leftrightarrow \nu)\biggr] \nn
&=& \Fmn(x) \,+\, \O(a^4) \, .
\eea
}
Using a suitable discretization of the second-order derivatives acting
on the field strength, we can therefore define a more 
continuum-like lattice field strength via
\bea\label{impcloverrep}
 \Fcmn(x) &\equiv&  
    {5\over 3}  \, \Fmn^{(cl)}(x) \; - \; {1\over 6} \, \biggl[
    U_\mu(x) \Fmn^{(cl)}(x+\mu)U_{-\mu}(x+\mu) \+ \nn
& & \qquad\qquad\qquad\quad \, \,
 U_{-\mu}(x) \Fmn^{(cl)}(x-\mu)U_{\mu}(x-\mu)  \-
 (\mu\leftrightarrow \nu)\biggr] \nn
&=& \Fmn(x) \,+\, \O(a^4) \, .
\eea

\vskip 3mm

\section{Dispersion Relation of the D234 Actions}

In terms of $\pbar$ and $\phat$ introduced in appendix~B let us define
\bea
\tilde{p}_\mu &\equiv& \pbar_\mu ( 1 + b_\mu a_\mu^2 \phat^2_\mu ) \nn
\tilde{m}(p)  &\equiv& m_0 + {1\over 2} r a_0 \sum_\mu \phat_\mu^2 + 
                                \sum_\mu c_\mu a_\mu^3 \phat_\mu^4 \nn
m_0           &\equiv& m_c ( 1 + {1\over 2} r a_0 m_c) \, .
\eea
The inverse of the free propagator of the general D234 action~\eqn{MDiiii}
is then simply $i\tilde{p}\sslash + \tilde{m}(p)$ in momentum space, 
implying the dispersion relation
\bea
        \tilde{p}^2 + \tilde{m}(p)^2  \,=\, 0 \, .
\eea
As a dimensionless measure of the energy it is useful to introduce
\bea
        y ~\equiv~ y(E) ~\equiv~ 
        -a_0^2 \phat_0^2 ~=~ 4 \sinh^2({a_0 E\over 2}) \, ,
\eea
so that
\bea
        -a_0^2 \pbar_0^2     &=& y~(1+{y\over 4}) \nn
        -a_0^2 \tilde{p}_0^2 &=& y~(1+{y\over 4})~(1-b_0 y)^2 ~.
\eea
Expressed in terms of $y$ the dispersion relation for the
energy $E=E({\bf p})$ therefore reads
\bea\label{ydispreln}
  y~(1+{y\over 4})~(1-b_0 ~y)^2 ~=~ a_0^2 \tilde{{\bf p}}^2 +
      \bigl ( \mu({\bf p}) - {r\over 2} ~y + c_0 ~y^2 \bigr )^2 ~,
\eea
where
\bea
\mu({\bf p}) ~\equiv~ a_0 m_0 + {1\over 2} r a_0^2 \sum_i \phat_i^2 +
                          a_0 \sum_i c_i a_i^3 \phat_i^4 ~.
\eea
Note that the use of $y$ factors out the particle anti-particle symmetry
$E\leftrightarrow -E$, 
since $y$ is invariant under it. The dispersion relation is generically
a quartic equation for $y$ with real coefficients. This implies that generically
there will be four solutions for $\pm E$, which, at a given ${\bf p}$,
are either real, come in complex conjugate pairs, or have imaginary part $\pi/a_0$.

Let us now discuss when the
quartic reduces to some lower order polynomial. We see immediately that the
quartic term cancels when $c_0 = {1\over 2} b_0$. Assuming this holds, the
cubic term cancels if $b_0 \seq 0$ or $r \seq 1-2b_0$. The remaining quadratic
equation reads
\bea\label{yquadr}
 -y^2 \, \Big[b_0 (2+\mu({\bf p})) + {1\over 4}(r^2 -1) \Big] +
 y \, \Big[ 1 + r \mu({\bf p}) \Big] ~=~ \mu({\bf p})^2 + a_0^2 
 \tilde{{\bf p}}^2 ~.
\eea
Note that the only way to obtain just one branch is to choose 
$r\seq 1, b_0 \seq c_0 \seq 0$, which corresponds to the Wilson or SW 
action.

Modulo the slight changes necessary for the ghost-free D234gf action,
all dispersion-related results and plots presented throughout this paper can be 
obtained by analytical or numerical manipulation of eq.~\eqn{ydispreln}.
For completeness we include some more results concerning the small mass 
expansions
of $E(0)/m_c$ and $c({\bf p})$ defined by
$E({\bf p})^2 = E(0)^2 + {\bf p}^2 c({\bf p})^2$. Here $E({\bf p})$ is the
physical branch of the dispersion relation (cf.~sect.~\ref{sec:largem});
$c({\bf p})$ is defined only for (sufficiently small) momenta and masses,
where there is a well-defined physical branch.

The first few coefficients in the expansions
\bea\label{Ecexpn}
 {E(0)\over m_c} &=& 1 \+ \sum_{n=1}^{\infty} \, E_n \, a_0^n m_c^n \nn
      c(0)       &=& 1 \+ \sum_{n=1}^{\infty} \, C_n \, a_0^n m_c^n
\eea
are presented in tables~\ref{tab:Eexpn} and~\ref{tab:cexpn} 
for various actions discussed in sect.~\ref{sec:Impr}.
These results are easily derived by an iterative series expansion of 
eq.~\eqn{ydispreln}, using a symbolic manipulation program.
Note that $E(0)/m_c$ and $c(0)$ are independent of 
the spatial $\De_i^2$ terms. For the D34 action we therefore only
have to specify $c_0$ in tables~\ref{tab:Eexpn} and~\ref{tab:cexpn}. 
This also implies that D234i(${2\over 3}$)
and the isotropic D234 action have the same $E(0)/m_c$ and $c(0)$
expansion coefficients,
cf.~tables~\ref{tab:Eexpn} and~\ref{tab:cexpn}.

For the SW action with $r\seq 1$ the exact mass $E(0)$ is given by the 
simple formula
\beq
 a_0 E(0)  ~=~ \ln\Big[1+a_0 m_c(1+{1\over 2} a_0 m_c)\Big] ~.
\eeq
We leave it as an exercise to the reader to derive analytical results for various
other special cases.

\begin{table}[tb] \centering
\begin{tabular}{ | l | r | r | r | r | r | r | }
\hline
Action      &  $E_1$  & $E_2$  &  $E_3$  &  $E_4$~  &  $E_5$~  &  
   $E_6$~\\ \hline
SW\rule[-2mm]{0mm}{7mm} & 0 &$-{1\o 6}$&${1\o 8}$ & $-{1\o 20}$ &
   $0$&${1\o 56}$\\
D34($c_0\!=\!{1\o 6}$)\rule{0mm}{5mm} & 0 & 0 &${1\o 6}$ & ${1\o 30}$ &
   ${1\o 36}$&${29\o 252}$\\
D234$\,\equiv\,$W2\rule{0mm}{5mm} & 0 & 0 &${1\o 18}$ & $-{1\o 270}$ &
   ${5\o 324}$&${41\o 6804}$\\
D234i(${2\o 3}$)\rule{0mm}{5mm} & 0 & 0 &${1\o 18}$ & $-{1\o 270}$ &
   ${5\o 324}$&${41\o 6804}$\\
D234(${2\o 3}$)\rule{0mm}{5mm} &  0  &  0  &  0 & ${1\o 30}$ &$-{5\o 216}$&
   ${115\o 6048}$\\
D234c(${2\o 3}$)\rule{0mm}{5mm} &  0  &  0  &  0 & ${1\o 30}$ &$-{1\o 54}$&
   ${37\o 2268}$\\[1mm]
\hline
\end{tabular}      
\vskip 2mm
\caption{Expansion coefficients of $E(0)/m_c$, eq.~\protect\eqn{Ecexpn}, 
         for various actions.
}
\label{tab:Eexpn}
\vskip 7mm
%
\begin{tabular}{ | l | r | r | r | r | r | r | }
\hline
Action      &  $C_1$  & $C_2$  &  $C_3$  &  $C_4$~  &  $C_5$~  &  
        $C_6$~\\ \hline
SW&\rule[-2mm]{0mm}{7mm} 0 &$-{1\o 3}$&${1\o 4}$ & $-{7\o 180}$ &
      $-{5\o 48}$&${1009\o 7560}$\\
D34($c_0\!=\!{1\o 6}$)\rule{0mm}{5mm}& 0 & 0 &${1\o 3}$ & ${1\o 10}$ &
       ${1\o 12}$&${22\o 63}$\\
D234$\,\equiv\,$W2 \rule{0mm}{5mm}& 0 & 0 &${1\o 9}$ & ${7\o 270}$ &
      ${7\o 324}$&${131\o 3402}$\\
D234i(${2\o 3}$)\rule{0mm}{5mm}& 0 & 0 &${1\o 9}$ & ${7\o 270}$ &
        ${7\o 324}$&${131\o 3402}$\\
D234(${2\o 3}$)\rule{0mm}{5mm} &  0  &  0  &  0 & ${1\o 10}$ &
        $-{5\o 72}$&${1135\o 18144}$\\
D234c(${2\o 3}$)\rule{0mm}{5mm} &  0  &  0  &  0 & ${1\o 10}$ &
         $-{1\o 18}$&${10\o 189}$\\[1mm]
\hline
\end{tabular}      
\vskip 2mm
\caption{Expansion coefficients of $c(0)$, eq.~\protect\eqn{Ecexpn}, 
         for various actions.
}
\label{tab:cexpn}
\vskip 4mm
\end{table}

The dispersion relations plotted throughout this paper apply, strictly
speaking, only to the on- but not off-shell improved actions obtained in the
third step of our procedure. The reader might wonder what happens to the 
dispersion relations, when, as instructed in step~3, we undo the change
of variable, leading to off-shell improved actions and fields. The off-shell
improved propagator in eq.~\eqn{corrprop} implies the dispersion relation
\beq
 \Tr(\Om(p)\Om^\star(p)) ~  \Tr(\Omb(p)\Omb^\star(p)) ~
 \Bigl( \tilde{p}^2 +\tilde{m}(p)^2 \Bigr) ~=~ 0 ~.
\eeq
The factors  from undoing field transformation give rise to additional 
high-lying ghosts, without affecting the other branches of the dispersion
relation. Considering, for simplicity, the SW case, where we can choose
$\Omb = \Om = 1 - {r a_0\o 4} (\del\slash - m_c)$, we have
\beq
{1\o 4} \, \Tr(\Om(p)\Om^\star(p)) ~=~  
{1\o 4} \, \Tr(\Omb(p)\Omb^\star(p)) ~=~
 \bigl(1 + {r a_0\o 4} m_c\bigr)^2 \+ \bigl({r a_0\o 4}\bigr)^2 \, \pbar^2 ~.
\eeq
For zero mass and momentum the extra ghosts are at energies
$a_0 E(0) = {\rm arcsinh}(4/r)$. If these ghosts are deemed to be not high
enough, one can push them up even further by modifying the field
transformation operators to read
\beq
  \Omb ~=~ \Om ~=~ \Bigl[ 1 - {r a_0\o 4 n} (\del\slash - m_c) \Bigr]^n
\eeq
for suitably large integer $n$. Obviously, similar remarks apply to the 
D234 actions.

\vskip 2mm

So far we have considered the dispersion relation in the absence of a 
background gauge field, finding that the coefficients of the 
scaling errors of our 
D234 actions are quite small. One might wonder if small 
coefficients prevail in the presence of a gauge field.

The short answer is yes, because, as for vanishing gauge field,
these errors ultimately arise from the truncation of the expansion of 
the the continuum $D_\mu$ in terms of lattice derivatives, which is a 
series with a finite radius of convergence and rapidly decreasing 
coefficients.

The longer answer goes as follows. When we discretize the continuum
action $M_\Om$ obtained by a change of variable
in  sect.~\ref{sec:Impr1},
the leading error is of even order, $a^2$, $a^4$, $\ldots$, and comes 
from the truncation of the $D\slash$ term; the truncation error of
$D\slash^2$ always comes with an additional factor of $a_0$.

The leading error of the general dispersion relation of a D234 action
(at least the D234c($r$) one, see below) is therefore simply that of
the naive improved fermion action $\delslc + m_c$. Remembering that
 $\del_{\! c\, \mu} = D_\mu - {1\o 30} a_\mu^4 D_\mu^5 + \O(a_\mu^6)\,\,$
we find the position-space dispersion relation
\beq
m_c^2 ~=~ \sum_\mu D_\mu^2 \+ {1\o 2}\si \! \cdot \! F
 \- {1\o 15}\sum_\mu a_\mu^4 D_\mu^6 \-
 {1\o 30}\sum_{\mu\nu}i \si_{\mu\nu} [D_\mu,a_\nu^4 D_\nu^5]  \+ \O(a^5) \, .
\eeq
The first two terms on the rhs give the continuum dispersion relation,
the rest the $\O(a^4)$ errors. Note that the commutator terms vanish
in the absence of a gauge field. As promised, the coefficients of the
error terms are small.

For the ``tuned'' D234 actions of sect.~\ref{sec:Impr}  there are additional
$a^3$ errors, \eg~there is a  $-{1\o 18} a_0^3 \De_0^2$~ 
correction to the action
for the D234i(${2\o 3}$) case. Again, the coefficients of these errors
are quite small.

For comparison, here is the dispersion relation of the SW action
\beq
m_c^2 ~=~  \sum_\mu D_\mu^2 \+ {1\o 2}\si \! \cdot \! F \+
  {1\o 3}\sum_\mu a_\mu^2 D_\mu^4 \+ 
  {1\o 6}\sum_{\mu\nu}i \si_{\mu\nu} [D_\mu,a_\nu^2 D_\nu^3]  \+ \O(a^3) \, .
\eeq

Note that using these expansions it is  trivial to calculate
the finite-mass corrections to various terms in the action. For example,
for the case of a small, constant, {\it magnetic} background field one 
finds that the energy of a ``D234c particle'' at zero velocity is:
\beq 
 E^2 ~=~ m_c^2 \, \Bigl( 1 + {1\o 15} a_0^4 m_c^4 \Bigr) \-
 {1\o 2} \, \si \!\cdot \! F \, \Bigl( 1 + {1\o 5} \, a_0^4 m_c^4 \Bigr ) 
 \+ \O(a^5) ~.
\eeq
The first term shows the correction corresponding to the leading coefficient
in table~\ref{tab:Eexpn}. The second shows that the correction to the
hyper-fine splitting, although not as small as for the mass, is still
small; on the one percent level as long as $a_0 m_c < 0.5$. This should
be compared with the SW case, where the analogous calculation gives
\beq 
 E^2 ~=~ m_c^2 \, \Bigl( 1 - {1\o 3} a_0^2 m_c^2 \Bigr) \-
 {1\o 2} \, \si \!\cdot \! F \, \Bigl( 1 - {2\o 3} \, a_0^2 m_c^2 \Bigr ) 
 \+ \O(a^3) ~,
\eeq
showing that $a_0 m_c$ must be smaller by almost a factor of 4 for
the hyper-fine splitting correction to be similarly small.

\section{Tadpole Calculus}

A large, apparently often dominant, part of the contributions of
standard lattice perturbation to a generic quantity
are unphysical, in the sense that
they are due to tadpole diagrams, which do not exist in
standard ways of performing
       continuum perturbation  theory.
 Lepage and Mackenzie~\cite{LM} proposed a simple method to improve 
lattice operators and actions, {\it i.e.}~make them more continuum-like.

At tree level, their prescription amounts to replacing each link field 
$U_\mu(x)$
in an operator by $U_\mu(x)/u_\mu$, where the number $0<u_\mu<1$ 
should be defined such that, roughly speaking, $1-u_\mu$ represents
the ``tadpole part'' of the link field.
At higher orders, one divides each link by the non-perturbatively measured
$u_\mu$, and multiplies by its perturbative expansion. In other words, one
reorganizes perturbation theory in a manner that sums up most of the
tadpole contributions to all orders, separately from the physical
contributions.  For isotropic actions
all $u_\mu$ are the same and a simple, gauge invariant prescription for
 the $u_\mu$ 
is given by  $u_0 = \langle {1\o N} \Tr~U_\Box \rangle^{1/4}$, 
where $U_\Box$ is the plaquette operator in a theory with SU($N$) gauge fields. 

This tadpole improvement prescription is the one used in the literature
up to now, and a version for anisotropic lattices was used for exploratory
simulations~\cite{LAT96,MorPea}. We are presently also exploring other
prescriptions~\cite{aniso}.

In the rest of this section we just assume
that the $u_\mu$ have been chosen by some scheme; their precise numerical
values are irrelevant.
For our fermionic actions the operators to be tadpole-improved are $\del_\mu$,
$\De_\mu$ and various products and sums thereof. 
Let $T(A)$ denote the tadpole-improved
version of some operator $A$. Note that $T$ is a linear operator, {\it i.e.}
\beq
 T(A+\la B) \, = \,  T(A) + \la T(B) ~~~~{\rm for~any}~ \la \in \IC ~. 
\eeq
Clearly,
\bea
 T(\del_\mu) &=& {\del_\mu\o u_\mu} \nn
 T(\De_\mu)  &=& {\De_\mu\o u_\mu} ~+~ 
                   {2\o a_\mu^2} ~\biggl( {1\o u_\mu^2} -1 \biggr) ~.
\eea
A subtlety arises for sufficiently complicated products of operators. Namely,
the improvement prescription requires one to expand out all products in terms
of $U_\mu(x)$'s and use $U_\mu(x) U_{-\mu}(x+\mu) \equiv 1$ for
``backtracking'' products of link fields. This implies that in
general
\beq
       T(A B) ~ \neq ~ T(A) ~ T(B) ~.
\eeq
In practice it will sometimes be more efficient to expand out a product of
operators in terms of link fields --- then no subtlety arises. In other cases,
however, it will  be more convenient or efficient to apply one tadpole-improved
operator after another. If we refer to the replacement of a product of
$\del_\mu$'s and $\De_\mu$'s by the product of their tadpole-improved versions
as ``naive tadpole improvement'', the simplest strategy in such a case is to 
first apply naive tadpole improvement and then correct the error by using the
following formulas:
\bea\label{TIeqs}
 T(\del_\mu \De_\mu) &=&  T(\del_\mu) \, T(\De_\mu) \nn
 T(\del_\mu^2)       &=&  T(\del_\mu)^2 \+ \de_\mu \nn
 T(\De_\mu^2)        &=&  T(\De_\mu)^2  \- {4\o a_\mu^2} \de_\mu \nn
 T(\del_\mu^3)       &=&  T(\del_\mu)^3 \+ {3\o 2}\de_\mu T(\del_\mu) \nn
 T(\De_\mu^3)        &=&  T(\De_\mu)^3 \- {6\o a_\mu^2} \de_\mu T(\De_\mu) \+
                                         {12\o a_\mu^4}\de_\mu 
\eea
where
\beq
 \de_\mu ~\equiv~ {1\o 2 a_\mu^2}\biggl( {1\o u_\mu^2} -1 \biggr ) ~.
\eeq
We also remind the reader of the identities~\eqn{delDeCom} and~\eqn{deldelexpn}.
Easy corrollaries of eqs.~\eqn{TIeqs} are
\bea
    T(\del\slash^2) &=& T(\del\slash)^2 \+ \sum_\mu \de_\mu \nn
    T(\del\slash^3) &=& T(\del\slash)^3 \+ (\sum_\mu \de_\mu) \, T(\del\slash) \+
                             {1\o 2} \sum_\mu \de_\mu \, T(\ga_\mu \del_\mu) 
\eea
which are useful for the tadpole improvement of
the field redefinition operators
$\Om$ and $\Omb$.


\newpage 

\begin{thebibliography}{99}

\bibitem{Gup} S.~Gupta, A.~Irb\"ack, F.~Karsch and B.~Petersson, 
           Phys.~Lett.~{\bf B242} (1990) 437.
\bibitem{Alf1} M.~Alford, W.~Dimm, G.P.~Lepage, G.~Hockney, P.B.~Mackenzie,
               Phys.~Lett.~{\bf B361} (1995) 87.
\bibitem{Sym} 
  K.~Symanzik, in: {\em Mathematical Problems in Theoretical Physics},
        R.~Schrader {\it et al.} (eds.), LNP~153, Springer, Berlin, 1982;
       in: {\em Recent Developments in Gauge Theories}, G.~'t~Hooft (ed.),
              Plenum Press, New York, 1980;
              Nucl. Phys. {\bf B226} (1983) 187, 205.
\bibitem{LW} M.~L\"uscher and P.~Weisz, Comm.~Math.~Phys. {\bf 97} (1985) 59,
             (E) {\bf 98} (1985) 433.
\bibitem{LWPLB} M.~L\"uscher and P.~Weisz, Phys.~Lett.~{\bf 158B} (1985) 250.
\bibitem{SW} B.~Sheikoleslami  and R.~Wohlert, Nucl.~Phys.~{\bf B259} (1985) 609.
\bibitem{LPCAC} M. L\"uscher, S. Sint, R. Sommer, P. Weisz, H. Wittig
         and U. Wolff, {\tt hep-lat/9608049} and references therein.
\bibitem{LM} G.P.~Lepage and P.B.~Mackenzie, Phys.~Rev. {\bf D48} (1993) 2250.
\bibitem{NRQCDccbar} C.T.H.~Davies, K.~Hornbostel, G.P.~Lepage, A.J.~Lidsey,
                   J.~Shigemitsu and J.~Sloan, Phys.~Rev. {\bf D52} (1995) 6519.
\bibitem{LAT95} M.~Alford, T.R.~Klassen, and G.P.~Lepage, 
      Nucl.~Phys. {\bf B} (Proc.~Suppl), {\bf 47} (1996) 370.
\bibitem{LAT96} M.~Alford, T.R.~Klassen, and G.P.~Lepage, 
 \verb:hep-lat/9608113:.
\bibitem{SCRI}  S. Collins, R. Edwards, U. Heller and J. Sloan,
               {\tt hep-lat/9608021}.
\bibitem{Borici} A.~Borici and Ph.~de Forcrand, {\tt hep-lat/9608105}.
\bibitem{Bock} W.~Bock, {\tt hep-lat/9608103}.
\bibitem{Bi} B. Beinlich, F. Karsch, and A. Peikert, {\tt hep-lat/9608141}.
\bibitem{Wil} K.G.~Wilson, in {\em New Phenomena in Subnuclear Physics}, Part~A,
     A.~Zichichi (ed.), Plenum Press, New York, p.~69, 1975.
\bibitem{Kar}   F.~Karsch, Nucl. Phys. {\bf B205} (1982) 285;
G.~Burgers, F.~Karsch, A.~Nakamura, and I.O.~Stamatescu, 
Nucl. Phys. {\bf B304} (1988) 587.
\bibitem{aniso} M.~Alford, T.R.~Klassen, G.P.~Lepage, C.~Morningstar, M.~Peardon,
  and H.~Trottier, {\em QCD on Anisotropic Lattices}, to appear.
\bibitem{Heatlie} G.~Heatlie, C.T.~Sachrajda, G.~Martinelli, C.~Pittori
                  and G.C.~Rossi, Nucl. Phys. {\bf B352} (1991) 266.
\bibitem{NRQCD} G.P.~Lepage, L.~Magnea, C.~Nakhleh, U.~Magnea and K.~Hornbostel,
                Phys. Rev. {\bf D43} (1992) 4052.
\bibitem{FNAL} A.X.~El-Khadra, A.S.~Kronfeld and P.B.~Mackenzie,
    {\tt hep-lat/9604004}.
\bibitem{AKL} M.~Alford, T.R.~Klassen and G.P.~Lepage, work in progress.
\bibitem{MorPea} C.~Morningstar and M.~Peardon,
                {\tt hep-lat/9608050}, {\tt hep-lat/9608019}.
\bibitem{Melb} G.P.~Lepage,  Nucl.~Phys. {\bf B} (Proc.~Suppl), 
      {\bf 47} (1996) 3.
\bibitem{Schl} G.P.~Lepage,  Schladming Winter School lectures, 
       {\tt hep-lat/9607076}.
%
\hide{
\bibitem{Weisz} P.~Weisz, Nucl.~Phys.~{\bf B212} (1983) 1;
                P.~Weisz and R.~Wohlert, Phys.~{\bf B236} (1984) 397.
\bibitem{Curci} G.~Curci, P.~Menotti and G.~Paffuti, Phys.~Lett.~{\bf 130B} 
(1983) 205, (E) {\bf 135B} (1984) 516.
\bibitem{WilGlue} K.~Wilson, Phys.~Rev.~{\bf D10} (1974) 2445.
}
%
\bibitem{NiNi} H.B.~Nielsen and M.~Ninomiya, Nucl. Phys. {\bf B185} (1981) 20;
                                                       {\bf B193} (1981) 173.
\bibitem{NN} R.~Narayanan and H.~Neuberger, Nucl. Phys. {\bf B443} (1995) 305.
\bibitem{KarSmit} L.H.~Karsten and J.~Smit, Nucl.~Phys.~{\bf B183} (1981) 103.
\hide{
\bibitem{KS} J.~Kogut and L.~Susskind, Phys.~Rev.~{\bf D11} (1975) 395;
   T.~Banks, J.~Kogut and L.~Susskind, Phys.~Rev.~{\bf D13} (1976) 1043;
                          L.~Susskind, Phys.~Rev.~{\bf D16} (1977) 3031.
}
\bibitem{EguKaw}  H.W.~Hamber and C.M.~Wu, Phys.~Lett. {\bf 133B} (1983) 351;
                  T.~Egushi and N.~Kawamoto, Nucl.~Phys.~{\bf B237} (1984) 609.
\bibitem{FW} H.R.~Fiebig and R.M.~Woloshyn,
   {\tt hep-lat/9603001};
    R.~Lewis and R.M.~Woloshyn, {\tt hep-lat/9610027};
   F.~Lee and D.~Leinweber, {\tt hep-lat/9606005}.
%
\bibitem{John}  We would like to thank J.~Sloan and R.~Edwards for discussions 
on this.

\end{thebibliography}
\end{document}